%% file: main.tex
\documentclass[useAMS,usenatbib, referee]{biom}

\usepackage{amsmath}
\usepackage{amsfonts}
\usepackage{algorithm}
\usepackage{algpseudocode}
\usepackage{algorithmicx}
\usepackage{gensymb}
\usepackage{comment}
\usepackage{graphicx}
\graphicspath{ {./} }
\usepackage{xcolor}
\usepackage{enumitem}

\algrenewcommand\algorithmicrequire{\textbf{Input:}}
\algrenewcommand\algorithmicensure{\textbf{Output:}}
\def\bSig\mathbf{\Sigma}

\newcommand{\tr}{\mbox{tr}}
\newcommand{\C}{\mathcal{C}}
\newcommand{\E}{\mathbb{E}}
\newcommand{\norm}[1]{\left\lVert#1\right\rVert}
\newcommand{\abs}[1]{\left\lvert#1\right\rvert}

\usepackage[colorlinks=true,linkcolor=blue,citecolor=blue,urlcolor=blue]{hyperref}
\usepackage{soul}

\title[Sparse network recovery from time course count data]{Sparse directed network recovery from multivariate count time series}

\author{Shirley Mathur$^{1,*}$%
, 
Nir Keret$^{2,*}$%
, and 
Ali Shojaie$^{1,2}$%
\\
$^{1}$Department of Statistics, University of Washington, Seattle, Washington, U.S.A. \\
$^{2}$Department of Biostatistics, University of Washington, Seattle, Washington, U.S.A. \\
$^{*}$These authors contributed equally to this work.
}

\begin{document}

\label{firstpage}

\begin{abstract}
Count time series arise naturally in many scientific applications. Examples include species abundances collected over time, and gene expression counts over pseudotime. Network inference from such data can yield important insight into underlying biological mechanisms, but is challenging because count observations are discrete, overdispersed, and temporally dependent. We propose a multivariate Poisson log-normal model in which observed counts arise from a Poisson distribution with rates determined by a latent vector autoregressive (VAR) process. The latent VAR transition matrix induces directed temporal dependencies among components and serves as the target network parameter. We establish identifiability of the model under suitable conditions and develop two procedures for sparse estimation of the VAR transition matrix. Our first estimator is a sparse method-of-moments estimator, for which we establish consistent network recovery and parameter estimation in high-dimensional settings. Our second estimator is an $\ell_1$-penalized variational inference estimator that is motivated by recent approaches for count data. 
We evaluate the proposed estimators through simulations under a range of sampling regimes. We further demonstrate the method on a macroinvertebrate abundance time series collected from European freshwater ecosystems, obtaining an interpretable ecological network.
\end{abstract}


\maketitle

\input{introduction.tex}

\input{methods.tex}

\input{simulations.tex}

\input{data_analysis.tex}

\input{discussion.tex}

\backmatter

\bibliographystyle{biom} 
\bibliography{refs}

\input{appendix.tex}

\label{lastpage}

\end{document}

%% file: introduction.tex
\section{Introduction}
\label{s:intro}

Multivariate count time series arise naturally in many biological applications, such as species abundances measured over time and longitudinal gene expression counts from measurements over time. A central goal is to recover an underlying network structure that captures interactions among system components (species in an environment, genes in a cell). However, unlike continuous-valued time series data, which have a well-developed modeling literature \citep{lutkepohl2005}, multivariate count time series have a less unified modeling framework. Their analysis is often complicated by over-dispersion in addition to the dependence among components and the serial dependence over time inherent to multivariate time series. Related work includes network inference for categorical time series \citep{tank2021convex} and for Poisson and Bernoulli autoregressive models \citep{hall2016inference}. Existing work provides several approaches to the problem, reviewed in \citet{Davis2021}. 
Early work developed integer-valued autoregressive models for univariate counts  \citep{mckenzie1985, alosh1987}, later  extended to the multivariate setting by \citet{latour1997}. However, these models do not admit an intuitive interpretation for network inference and are limited in their ability to capture general autocorrelation structures. Recent work addressed some of these limitations. For example, \citet{Yu2024} allows for flexible correlation in the noise component, but still does not allow one component to depend on past values of another. \citet{fokianos2020} developed multivariate count autoregressive models in which past observed counts directly influence future conditional means, with contemporaneous dependence captured through a copula. We instead treat counts as noisy observations of an underlying process whose dynamics define the target network. This separates temporal interactions in the latent log-rates from Poisson sampling variation.

To address the limitations of existing approaches, we adopt a multivariate Poisson log-normal (PLN) model \citep{aitchison1989}, in which observed counts follow a Poisson distribution with rates driven by a latent Gaussian vector autoregressive (VAR) process. This model naturally accommodates count-valued data, over-dispersion, and dependence both within and across time points. Moreover, the transition matrix of the latent VAR process encodes directed temporal dependencies among components corresponding to Granger causal interactions \citep{shojaie2022granger}, so its support can be interpreted as the underlying network structure.

Our proposed model is most closely related to \citet{Zhang2017}, who modeled multivariate Poisson counts using latent Gaussian VAR dynamics and estimated the parameters by Monte Carlo EM, with numerical studies primarily in low-dimensional settings. We focus on sparse recovery of the directed network encoded by the latent transition matrix, developing computationally scalable estimators with support-recovery guarantees. Our framework also accommodates covariates, multiple independent trajectories of unequal length, and $\ell_1$-penalized estimation for sparse recovery. For network inference, our approach is closer to \citet{chiquet2019}, who used a PLN model to infer an undirected network via sparsity in the latent precision matrix. Recent extensions of the PLN framework include classification, clustering, and dimension reduction (see \citet{chiquet2021} for an overview), as well as tree-structured PLN models for hierarchical count data \citep{chaussard2025}. \citet{saha2026} introduced the Poisson log-normal process, which extends the PLN framework by placing a Gaussian process prior on the Poisson log-rates for count prediction.

Many existing PLN methods rely on variational inference (VI), estimating model parameters by maximizing the evidence lower bound (ELBO) \citep{Blei2017}. More recent work has proposed likelihood-based alternatives, including maximum marginal likelihood estimation \citep{huang2025}, motivated by composite likelihood ideas \citep{cox2004, varin2008}. Relatedly, \citet{stoehr2026} combine composite likelihood with importance sampling. Full likelihood-based estimation requires integration over the latent trajectories, making computation challenging as the number of components and time points increases. Marginal likelihood approaches, such as \citet{huang2025}, reduce this burden by using low-dimensional marginal distributions, but still require numerical integration and optimization.

We propose a method-of-moments (MoM) estimator as well as its $l_1$-penalized version to encourage sparsity in high dimensions. Our penalized MoM estimator is related to \citet{duker2024},  who studied sparse latent VAR estimation for count time series. Their model generates counts by thresholding latent Gaussian variables, whereas ours uses a Poisson observation model conditional on the latent Gaussian states. Our proposed estimator scales well in high-dimensional settings and has theoretical guarantees on recovering the support of the network of interest under usual conditions for sparse estimation. We also develop a VI estimator and penalized VI estimator to compare our method to the prevailing estimation strategy used by PLN models in the literature. 

The main methodological contributions of this work are twofold. First, we formulate network recovery for multivariate count time series under a dynamic PLN model in which the support of the latent VAR transition matrix defines the target directed network. Second, we develop an estimation procedure for sparse network recovery along with theoretical guarantees. We illustrate the practical utility of our method using macroinvertebrate species abundance data from Danish freshwater environments in the TREAM (Time seRies of European freshwAter Macroinvertebrates) dataset \citep{tream2024}.

%% file: methods.tex
\section{Methodology}
\label{s:method}
\subsection{Model and network target}\label{sec:model_setup}

We consider $n$ independent realizations of a multivariate count time series, where $i$ indexes the realization and the timepoints for each realization are indexed along a grid of equally spaced timepoints $1, 2, \dots, T_i$, where $T_i$ is the latest timepoint at which the realization is observed. At time $t$, let $Y^{(i)}_t \in \mathbb{Z}_{\geq 0}^{d}$, denote the vector of counts for $d$ components, $X^{(i)}_t \in \mathbb{R}^p$, the vector of observed covariates (excluding an intercept term), and $O^{(i)}_t \in \mathbb{R}$ an offset. Let $Z^{(i)}_t \in \mathbb{R}^{d}$ denote the latent state for sample $i$ at time $t$ and let $u^{(i)}_t$ denote the corresponding noise term. Further, let $\boldsymbol{\Sigma} \in \mathbb{R}^{d \times d}$ be a positive definite covariance matrix, $A \in \mathbb{R}^{d \times d}$ the latent transition matrix, $\beta_0 \in \mathbb{R}^d$ the intercept vector, and $\beta \in \mathbb{R}^{p \times d}$ the matrix of regression coefficients. Since higher-order VAR processes can be written in ``companion form" as VAR(1) processes, for simplicity, we focus on a latent VAR(1) model \citep{lutkepohl2005}. Specifically, we assume
\begin{align}
&Z_t^{(i)} = AZ_{t-1}^{(i)} + u_t^{(i)}, \quad u_t^{(i)} \overset{\text{i.i.d.}}{\sim} N(\mathbf{0}, \mathbf{\Sigma}) \label{latent_dist}\\
\quad &Y^{(i)}_{tj} | Z^{(i)}_{tj}  \sim \text{Pois.}\left\{\exp\left(O^{(i)}_{t} + \beta_{0j} + X_{t}^{(i)\top}\beta_j + Z_{tj}^{(i)}\right)\right\} \label{obs_cov_dist}
\end{align}
We additionally assume that, conditional on the latent trajectory, covariates, and offsets, the counts are independent across time points and components. Suppressing the conditioning on covariates and offsets, the joint distribution of the latent states and observed counts factorizes as
\begin{align*}
p(y_{1:T},z_{1:T})&=p(z_1)\prod_{t=2}^Tp(z_t|z_{{t-1}})\prod_{t=1}^T\prod_{j=1}^dp(y_{tj}|z_{tj}) \, .
\end{align*}

We assume that the $n$ trajectories are independent and follow this model. Further, we assume all eigenvalues of $A$ have modulus less than $1$ and that the latent process is stationary. For each trajectory, the latent process is independent of the covariates and offsets. Letting $\boldsymbol{\Sigma}_Z \in \mathbb{R}^{d \times d}$ denote the stationary covariance matrix of $Z_t$, we have $\boldsymbol{\Sigma}_Z = A\boldsymbol{\Sigma}_ZA^{\top} + \boldsymbol{\Sigma}$ \citep{lutkepohl2005}.

Our target of inference is the network of Granger-causal interactions among the components of the latent process $Z_t$, as defined in \citet{shojaie2022granger}. In the proposed model, $A_{jk}$ represents the lagged effect of latent component $k$ on latent component $j$, and $A_{jk}\neq0$ corresponds to a directed edge from component $k$ to component $j$.

\subsection{Identifiability}
Under conditions on the design matrix of observed covariates and the covariance of the latent VAR process, the proposed model is identifiable. 

\begin{lemma} \label{lem:identification}
Let $N_{\mathrm{tot}} := \sum\limits_{i=1}^n T_i$, and let $\mathbf{X} \in \mathbb{R}^{N_{\mathrm{tot}} \times p}$ denote the matrix of observed covariates obtained by stacking $X_t^{(i)}$ over all samples and time points. If the augmented design $\bar{\mathbf{X}}=[\mathbf{1},\mathbf{X}]$ has full column rank, and $\boldsymbol{\Sigma}_Z$ is invertible, then the model is identifiable.
\end{lemma}

\begin{proof}
The proof is given in Appendix \hyperref[sec:a1]{A}.
\end{proof}

We next describe the two proposed estimation procedures for the model parameters. We start with the MoM estimator and then discuss the alternative VI estimator.

\subsection{Method of Moments Estimator}\label{sec:mom_estimator}
We develop a method-of-moments estimator for the model parameters $\theta := \left\{A, \beta_0, \beta, \boldsymbol{\Sigma}\right\}$, with primary interest in $A$, whose support determines the directed network. 

We first define some notation. Let $\norm{\cdot}_1$, $\norm{\cdot}_2$, and $\norm{\cdot}_\infty$ denote the usual $\ell_1$, Euclidean, and maximum norms of a vector. For a matrix $M$, 
$$
\norm{M}_{\max}=\max_{j,k}\abs{M_{jk}},\quad
\norm{M}_1=\max_k\sum_j\abs{M_{jk}},\quad
\norm{M}_\infty=\max_j\sum_k\abs{M_{jk}}.
$$
We write $\norm{M}_2$ for its spectral norm, and $\norm{M}_F$ for its Frobenius norm. 

We begin by estimating $\beta$. Let $\gamma \in \mathbb{R}^{(p + 1) \times d}$ be defined columnwise by $\gamma_j := [\,\beta_{0j} + \frac{1}{2}\boldsymbol{\Sigma}_{Z, jj}, \, \beta_{1j}, \, \ldots ,\,  \beta_{pj} \,]^{\top} \in \mathbb{R}^{p + 1}$, and index the rows of $\gamma$ from $0$ to $p$. Since $\gamma_{1:p}=\beta$, estimating $\gamma$ yields an estimator of $\beta$. Moreover, $\mathbb{E}\left[Y_{tj}^{(i)} \mid {X}_{t}^{(i)}, {O}_t^{(i)}\right] = \exp\left\{O_{t}^{(i)} + \gamma_{0j} + X_t^{(i)\top}\gamma_{1:p,j} \right\}$, so $\gamma$ can be estimated  from the first moment by fitting a Poisson regression. Concretely, we define the objective $L_j(g)$ for $g = \left[g_0 \ g_{1:p} \right]^{\top} \in \mathbb{R}^{p+1}$ as
\begin{equation*}
L_j(g) = -\frac{1}{N_{tot}}\sum\limits_{i=1}^n \sum\limits_{t=1}^{T_i}\left\{ Y_{tj}^{(i)}\left(O_{t}^{(i)} + g_0 +  X_{t}^{(i)\top}g_{1:p}  \right) - \exp\left(O_{t}^{(i)} + g_0 +  X_{t}^{(i)\top}g_{1:p}  \right)\right\},
\end{equation*}
and obtain estimates for $\gamma_j$ ($j=1,\ldots,d$) as
$\hat{\gamma}_j = \arg\min_{g} L_j(g)$.

It remains to estimate $\beta_0$. By the definition of $\gamma$, we have $\hat{\beta}_0 = \hat{\gamma}_0 - \frac{1}{2}\mathrm{diag}\left(\boldsymbol{\hat{\Sigma}}_Z \right)$, where $\mathrm{diag}(\cdot)$ returns the diagonal of a matrix as a vector. We therefore estimate $\boldsymbol{\Sigma}_Z$ from the second moments $\mathbb{E}\left[Y^{(i)}_{tj}Y^{(i)}_{tk} \mid X_t^{(i)}, O_t^{(i)}\right]$, for all $j,k = 1, \dots, d$. The explicit expressions for estimating $\boldsymbol{\Sigma}_Z$ are 
\begin{align*}
\exp\left\{\hat{\Sigma}_{Z,jj} \right\} = \frac{1}{\sum_{i=1}^nT_i}\sum\limits_{i=1}^{n}\sum\limits_{t=1}^{T_i}\frac{Y_{tj}^{(i)2} - Y_{tj}^{(i)}}{\exp\left\{2\left(O_{t}^{(i)} + \hat{\gamma}_{0j} + X_t^{(i)\top}\hat{\gamma}_{1:p,j}\right)\right\}} \, , \\
\exp\left\{\hat{\Sigma}_{Z,jk} \right\} = \frac{1}{\sum_{i=1}^nT_i}\sum\limits_{i=1}^{n}\sum\limits_{t=1}^{T_i}\frac{Y_{tj}^{(i)}Y_{tk}^{(i)}}{\exp\left\{2O_{t}^{(i)} + \hat{\gamma}_{0j} + \hat{\gamma}_{0k} + X_t^{(i)\top}\left(\hat{\gamma}_{1:p,j} + \hat{\gamma}_{1:p,k}\right)\right\}} \, .
\end{align*}

Having estimated $\beta$, we proceed to estimating $A$. Let $P \in \mathbb{R}^{d \times d}$ be a matrix that captures the lag-one cross-moments,
\begin{equation*}
P_{jk} := \log\left(\frac{\mathbb{E}\left[Y^{(i)}_{tj}Y^{(i)}_{t-1,k} \mid X^{(i)}_t, X^{(i)}_{t-1}, O^{(i)}_t, O^{(i)}_{t-1}\right]}{\mathbb{E}[Y^{(i)}_{tj} \mid X^{(i)}_t, O^{(i)}_t]\mathbb{E}[Y^{(i)}_{t-1,k} \mid X^{(i)}_{t-1}, O^{(i)}_{t-1}]}\right) \, .
\end{equation*}

In our model, the lag-one moments imply $P = A\boldsymbol{\Sigma}_Z$. Hence, we propose first estimating $P$ and then setting $\hat{A} = \hat{P}\hat{\boldsymbol{\Sigma}}^{-1}_Z$. The explicit estimating equation for each element of $P$ is
\begin{align*}
\exp\left\{\hat{P}_{jk} \right\} = \frac{1}{\sum_{i=1}^n \left(T_i - 1\right)}\sum\limits_{i=1}^{n}\sum\limits_{t=2}^{T_i} \frac{Y_{tj}^{(i)}Y_{t-1,k}^{(i)}}{\exp\left\{O_{t}^{(i)} + O_{t-1}^{(i)} + \hat{\gamma}_{0j} + \hat{\gamma}_{0k} + X_t^{(i)\top}\hat{\gamma}_{1:p,j} + X_{t-1}^{(i)\top}\hat{\gamma}_{1:p,k} \right\}} \, .
\end{align*}
Finally, leveraging stationarity,  we estimate $\boldsymbol{\Sigma}$ by the plug-in estimator $\hat{\boldsymbol{\Sigma}} = \hat{\boldsymbol{\Sigma}}_Z - \hat{A}\hat{\boldsymbol{\Sigma}}_Z\hat{A}^{\top}$. 

We show that under the conditions  stated in Appendix~\hyperref[sec:a2]{B}, the components of  $\hat{\gamma}$, $\hat{\boldsymbol{\Sigma}}_Z$ and $\hat{P}$ converge to their target values. 

\begin{lemma}
\label{lem:mom-rates}
Under Assumptions~\ref{as:trajectories}--\ref{as:dimension_growth}, as $n \to \infty$, 
$$
\max_{1\leq j\leq d}\norm{\hat{\gamma}_j-\gamma_j}_2= O_p\left(\sqrt{\frac{\log d}{n}}\right)
$$
and,
$$\norm{\hat{\Sigma}_Z - \Sigma_Z}_{\max} = O_p\left(\sqrt{\frac{\log d}{n}}\right), \qquad 
\norm{\hat{P} - P}_{\max} = O_p\left(\sqrt{\frac{\log d}{n}}\right) \, .
$$
\end{lemma}

Assumptions ~\ref{as:trajectories}--\ref{as:stationarity} state that the number of time-points for each realization of the time series is fixed and does not grow as $n$ grows, and the assumed model defined in Section \ref{sec:model_setup}. The bounded-length condition reflects the longitudinal-data setting we consider, where each subject has a limited number of serial observations. Thus, our asymptotic results concern increasing $n$, allowing $d$ to grow while the trajectory lengths remain uniformly bounded. Assumptions~\ref{as:boundedness}--\ref{as:gamma_Hessian} impose moment and compactness conditions on the covariates and parameters, bounded offsets, and uniform positive curvature of the population Poisson criterion. These conditions allow us to establish consistency of $\hat{\gamma}_j$ uniformly over $j$. Assumption~\ref{as:dimension_growth} bounds the entries of $\boldsymbol{\Sigma}_Z$ and $P$ and restricts the growth of $d$ relative to $n$.

\subsection{Sparse MoM Estimator}\label{sec:mom_estimator_pen}
To facilitate estimation in high dimensions, we next develop a procedure to estimate sparse $A$ matrices. 

First, note that our MoM estimator  for $A$ in Section~\ref{sec:mom_estimator} is equivalent to solving the optimization problem
$$
\hat{A} = \underset{A}{\arg\min} \norm{A\hat{\boldsymbol{\Sigma}}_Z - \hat{P}}^2_F \ .
$$
Noticing that the above problem separates over rows, and using the symmetry of $\hat{\Sigma}_Z$, each row $a^{(j)}$ for all $j \in \left\{1, 2, \dots, d \right\}$ can be estimated as 
$$
\hat{a}^{(j)} = \underset{a^{(j)} \in \mathbb{R}^{d}}{\arg\min} \norm{\hat{\Sigma}_Za^{(j)} - \hat{p}^{(j)}}_2^2 \ ,
$$
where $\hat{p}^{(j)}$ denotes the $j$-th row of $\hat{P}$.

In light of framing the estimation of $A$ as a regression problem, we propose a two-step estimation procedure for estimation of a sparse $A$ similar to \citet{belloni2013}. 
More specifically, we first use a variable selection procedure to recover the support for each row of $A$. Then, we refit each row on its selected support to obtain the final estimate of $A$. When $A$ is not too large, the first step can be accomplished via an exhaustive search over all possible supports and selecting the one that minimizes an information criterion, such as BIC, for the above loss. When it is computationally infeasible to search over all potential supports for $A$, we perform the variable selection step through fitting an initial sparse MoM estimator with a weighted-$\ell_1$ penalty on $A$. Specifically, after estimating $\beta, \boldsymbol{\Sigma}_Z,$ and $P$ as above, we set the weights as $\hat{w}_{jk}=\hat{\Sigma}^{1/2}_{Z,kk}/\hat{\Sigma}^{1/2}_{Z,jj}$. These weights make penalization comparable across the marginal scales of the latent variables. We find the penalized estimate for each row of $A$ with
\begin{equation} \label{eq:weighted-row-mom}
\hat{a}_{\lambda_j}^{(j)} = \underset{a^{(j)} \in \mathbb{R}^{d}}{\arg\min} \norm{\hat{\Sigma}_Za^{(j)} - \hat{p}^{(j)}}_2^2 + \lambda_{j}\sum\limits_{k=1}^d \hat{w}_{jk}\abs{a^{(j)}_k} \, .
\end{equation}

The full penalized estimate for $A$ is then written as $\hat{A}_{\ell_1} = \begin{bmatrix} \hat{a}_{\lambda_1}^{(1)} & \cdots & \hat{a}_{\lambda_d}^{(d)} \end{bmatrix}^{\top} \in \mathbb{R}^{d \times d}$.

\begin{theorem}\label{thm:support-recovery}
Under Assumptions~\ref{as:trajectories}--\ref{lambda_selection} in Appendix~\ref{sec:a2}, the optimization problem \eqref{eq:weighted-row-mom} has a unique minimizer for every $j=1,\ldots,d$ with probability tending to one. The resulting estimators satisfy
$$
\Pr\left\{\operatorname{supp}(\hat{A}_{\ell_1})=\operatorname{supp}(A),
\ \operatorname{sign}(\hat{A}_{\ell_1})=\operatorname{sign}(A)\right\}\longrightarrow1.
$$
\end{theorem}

Assumptions~\ref{as:trajectories}--\ref{as:dimension_growth} establish the preliminary rates in Lemma~\ref{lem:mom-rates}. Assumption~\ref{sigmaZ_cov_bound} bounds the sum of absolute covariances in each column of $\boldsymbol{\Sigma}_Z$ uniformly as $d$ grows. Assumptions~\ref{eigen_min}--\ref{pop_irrepresentability} require well-conditioned active Gram matrices and a weighted irrepresentability condition, as is standard in signed support recovery for $\ell_1$-penalized regression \citep{wainwright09}. Assumption~\ref{as:weight_variances} keeps the marginal latent variances uniformly bounded away from zero, allowing consistent estimation of the ratio weights in \eqref{eq:weighted-row-mom}. Assumption~\ref{A_max_bound} bounds the entries of $A$, while Assumption~\ref{lambda_selection} specifies penalty sequences that dominate the moment-estimation error while remaining small relative to the weakest nonzero signal.

The optimization problem for the initial penalized estimator in \eqref{eq:weighted-row-mom} is solved using the fast iterative shrinkage-thresholding algorithm (FISTA) \citep{beck2009}. The selection of an appropriate tuning parameter for estimating equations in high-dimensional settings is generally a difficult problem, as discussed in \citet{xia2026} and \citet{neykov2018}. We propose to select the tuning parameters by using a BIC-like criterion. We first fit the penalized estimator over a grid of candidate $\lambda_j$ values, and for each selected $\lambda_j$, refit on the selected support with
\begin{align*}
\label{eq:mom-pen-refit}
\hat{a}^{(j)}_{\mathrm{refit}, \lambda_j} &:= \underset{a^{(j)}}{\arg\min} \norm{\hat{\boldsymbol{\Sigma}}_Za^{(j)} - \hat{p}^{(j)}}^2_2 , &\textrm{ s.t.  } \mathrm{supp}(a^{(j)}) = \mathrm{supp}\left(\hat{a}^{(j)}_{\lambda_j}\right), 
\end{align*}
where $\mathrm{supp}\left(a^{(j)}\right)$ is the support of $a^{(j)}$. 
Then, we select $\lambda_j$ by minimizing the following BIC-like criterion, 
\begin{equation*}
\text{BIC}_{\text{MoM},j}(\lambda_j) := n\norm{\hat{\boldsymbol{\Sigma}}_Z\hat{a}^{(j)}_{\mathrm{refit}, \lambda_j} - \hat{p}^{(j)}}^2_2 + \log \left(n\right)\left|\mathrm{supp}\left(\hat{a}^{(j)}_{\mathrm{ refit}, \lambda_j}\right)\right| \ .
\end{equation*}

The final estimator for $A$ is then obtained as $\hat{A}_{\mathrm{sparse}} = \begin{bmatrix} \hat{a}_{\mathrm{refit}, \lambda^*_1}^{(1)} & \cdots & \hat{a}_{\mathrm{refit}, \lambda^*_d}^{(d)} \end{bmatrix}^{\top} \in \mathbb{R}^{d \times d}$, where $\lambda_j^*$ denotes the selected $\lambda_j$ for the $j$-th row. For fixed $d$ and fixed model parameters, the BIC-type criterion consistently selects the true support, provided $\Sigma_Z$ is invertible and the candidate supports include the true support with probability tending to one. Indeed, Lemma~\ref{lem:mom-rates} gives $n^{-1/2}$ convergence rates for $\hat{\Sigma}_Z$ and $\hat{P}$. The refitted loss on the true support is therefore $O_p(n^{-1})$. Adding unnecessary coefficients can reduce the loss by at most this amount, giving an $O_p(1)$ improvement in the $n$-scaled loss, whereas each additional coefficient incurs a $\log(n)$ penalty. Conversely, invertibility of $\Sigma_Z$ implies that every support omitting a true nonzero coefficient has strictly positive population loss. Its $n$-scaled loss therefore grows at order $n$, dominating any reduction in the penalty, which is of order $\log(n)$. Since there are finitely many supports, the true supports are selected simultaneously for all rows with probability tending to one. When $d$ grows with $n$, controlling the increasing number of candidate supports may require an additional dimension-dependent complexity penalty, as in the extended BIC (EBIC).

Our final sparse estimate $\hat{A}_{\mathrm{sparse}}$ is a consistent estimator of $A$ as a consequence of Theorem~\ref{thm:support-recovery}. 

\begin{corollary}\label{cor:coefficient-consistency}
Let $s_n=\max_{1\leq j\leq d}|\mathrm{supp}\left(a^{(j)} \right)|$ and $\rho_n = r_n+dr_n^2$, where $r_n = \sqrt{\frac{\log d}{n}}$. Under the assumptions of Theorem~\ref{thm:support-recovery},
$$
\norm{\hat{A}_{\ell_1}-A}_{\max} = O_p\left(s_n^{3/2}\rho_n+\sqrt{s_n}\max_j\lambda_j\right)=o_p(1).
$$
Let $\hat{A}_{\mathrm{sparse}}$ be obtained by minimizing the
unpenalized rowwise loss on the estimated supports
$\hat{S}^{(j)}=\operatorname{supp}(\hat{a}_{\lambda^*_j}^{(j)})$, with
coefficients outside these supports set to zero. If the selected supports equal the true supports with probability tending to one, then
$$
\norm{\hat{A}_{\mathrm{sparse}}-A}_{\max}
=O_p(s_n^{3/2}\rho_n)=o_p(1).
$$
\end{corollary}

\subsection{MoM Estimator for partially observed time series}
Our MoM estimator can also be applied in settings where there are gaps in the observed time series, i.e. some observations are missing. Let $R^{(i)}_{tj} \in \left\{0,1 \right\}$ be the missingness indicator for observation $Y^{(i)}_{tj}$ so that $R^{(i)}_{tj} = 1$ when $Y^{(i)}_{tj}$ is observed and $R^{(i)}_{tj} = 0$ otherwise, and collect the missingness indicators and observations for all time series realizations, timepoints, and categories as $\mathbf{R}, \mathbf{Y}$, respectively. Then, assuming that all covariates are observed and that missingness is random conditional on the covariates and offset---i.e., $\mathbf{R} \perp \mathbf{Y} \mid \mathbf{X}, \mathbf{O}$---we can still obtain unbiased estimating equations for $\theta$. First, the objective for estimating each $\gamma_j$ is now written as
\begin{equation*}
L_j(g) = -\frac{1}{\sum\limits_{i=1}^n\sum\limits_{t=1}^{T_i} R_{tj}^{(i)}}\left\{\sum\limits_{i=1}^n \sum\limits_{t=1}^{T_i} R_{tj}^{(i)}\left[Y_{tj}^{(i)}\left(O_{t}^{(i)} + g_0 +  X_{t}^{(i)\top}g_{1:p}  \right) - \exp\left(O_{t}^{(i)} + g_0 +  X_{t}^{(i)\top}g_{1:p}  \right)\right]\right\} \, .
\end{equation*}

Then, the solutions of the estimating equation for $\boldsymbol{\Sigma}_Z$ are
\begin{align*}
\exp\left\{\hat{\Sigma}_{Z,jj} \right\} = \frac{1}{\sum_{i=1}^n\sum\limits_{t=1}^{T_i} R_{tj}^{(i)}}\sum\limits_{i=1}^{n}\sum\limits_{t=1}^{T_i}\frac{R_{tj}^{(i)}\left(Y_{tj}^{(i)2} - Y_{tj}^{(i)}\right)}{\exp\left\{2\left(O_{t}^{(i)} + \hat{\gamma}_{0j} + X_t^{(i)\top}\hat{\gamma}_{1:p,j}\right)\right\}} \, , \\
\exp\left\{\hat{\Sigma}_{Z,jk} \right\} = \frac{1}{\sum_{i=1}^n \sum\limits_{t=1}^{T_i} R_{tj}^{(i)}R_{tk}^{(i)}}\sum\limits_{i=1}^{n}\sum\limits_{t=1}^{T_i}\frac{R_{tj}^{(i)}R_{tk}^{(i)}Y_{tj}^{(i)}Y_{tk}^{(i)}}{\exp\left\{2O_{t}^{(i)} + \hat{\gamma}_{0j} + \hat{\gamma}_{0k} + X_t^{(i)\top}\left(\hat{\gamma}_{1:p,j} + \hat{\gamma}_{1:p,k}\right)\right\}} \, .
\end{align*}

Next, the solution of the estimating equation for each element of $P$ is
\begin{align*}
\exp\left\{\hat{P}_{jk} \right\} = \frac{1}{\sum_{i=1}^n \sum\limits_{t=2}^{T_i} R^{(i)}_{tj}R^{(i)}_{t-1,k}}\sum\limits_{i=1}^{n}\sum\limits_{t=2}^{T_i} \frac{R^{(i)}_{tj}R^{(i)}_{t-1,k}Y_{tj}^{(i)}Y_{t-1,k}^{(i)}}{\exp\left\{O_{t}^{(i)} + O_{t-1}^{(i)} + \hat{\gamma}_{0j} + \hat{\gamma}_{0k} + X_t^{(i)\top}\hat{\gamma}_{1:p,j} + X_{t-1}^{(i)\top}\hat{\gamma}_{1:p,k} \right\}} \, .
\end{align*}

The estimator for $A$ is obtained as outlined earlier in Sections~\ref{sec:mom_estimator} and \ref{sec:mom_estimator_pen} for the standard and sparse MoM estimators, respectively, and the estimator for $\boldsymbol{\Sigma}$ is obtained as specified in Section~\ref{sec:mom_estimator}.

\subsection{Variational Inference Estimator}
An alternative to the proposed method-of-moments estimator is a likelihood-based estimation strategy. However, exact likelihood-based estimation for our PLN model is challenging because the observed-data likelihood, $ p_{\theta}(y)$, requires integrating over the latent states, which is not tractable. A standard approach for latent-variable models is the EM algorithm \citep{dempster1977} but here the required conditional moments are not available in closed form, and Markov-Chain Monte Carlo-based approximations would be computationally prohibitive in high dimensions. As a result, variational inference (VI) approaches, wherein parameters are estimated by maximizing the evidence lower bound (ELBO), have become prevalent in the PLN literature \citep{chiquet2019,chiquet2021}. In this section, we develop a penalized VI estimator for our PLN model. 

Under stationarity, the initial latent density 
$p_{\theta}(z_1)$ is Gaussian with covariance $\boldsymbol{\Sigma}_Z$. If one were to derive the ELBO from the full likelihood, the resulting objective would involve $\log\det(\boldsymbol{\Sigma}_Z)$  and $\boldsymbol{\Sigma}_Z^{-1}$. Because $\boldsymbol{\Sigma}_Z$ is defined implicitly by $A$ and $\boldsymbol{\Sigma}$, this would complicate optimization. We therefore base the ELBO on the likelihood conditional on $Z_1 \in \mathbb{R}^{n \times d}$, the collection of the latent states at the first time point across samples, and treat $Z_1$ as a set of fixed nuisance parameters to be estimated. 
Let $q_{\Psi}(Z)$ denote a variational approximation to the conditional
posterior $p_{\theta}(z\mid y,x,o,z_1)$ of the latent states at times $t\ge 2$, where $\Psi$ collects the variational means and variances together with the unknown initial-state values. We use a mean-field family for the variational approximation, defined as
\begin{equation*}
    Q := \left\{q : q_{\Psi}(z) = \prod_{i=1}^n\prod_{t=2}^{T_i}\prod_{j=1}^d \phi\left(z^{(i)}_{tj}; m^{(i)}_{tj}, s^{(i)}_{tj}\right) \right\}\, ,
\end{equation*}
where $\phi(\cdot;\mu,\sigma^2)$ denotes the univariate Gaussian density with mean $\mu$ and variance $\sigma^2$. For $t\geq2$, $m^{(i)}_{tj}$ and  $s^{(i)}_{tj}$ are the variational mean and variance  of $Z^{(i)}_{tj}$, respectively. Because we condition on $Z_1$, the initial states are treated as fixed, with their unknown values estimated through $m^{(i)}_{1j}$. We set $s^{(i)}_{1j}=0$ as a notational convention in the ELBO below. 

With this variational approximation, the form of the conditional ELBO, up to parameter-independent additive terms, is 
\begin{align*}
\text{ELBO}(y; \Psi, \theta) &= \sum\limits_{i=1}^n \sum\limits_{t=1}^{T_i} \left[ y_t^{(i)\top}\left({m}_{t}^{(i)} + \mu_{t}^{(i)} \right) - \mathbf{1}_d^{\top}\exp\left\{\mu_{t}^{(i)} + {m}_{t}^{(i)} + \frac{1}{2}{s}_{t}^{(i)} \right\}  \right] \\
&\quad -\frac{1}{2}\sum\limits_{i=1}^n (T_i - 1)\log\det(\boldsymbol{\Sigma})  +\frac{1}{2}\sum\limits_{i=1}^n\sum\limits_{t=2}^{T_i} \sum\limits_{j=1}^d\log(s_{tj}^{(i)}) \\
&\quad - \frac{1}{2}\sum\limits_{i=1}^n\sum\limits_{t=1}^{T_i -1} \tr\left\{\left[\left(m_{t+1}^{(i)} - Am_t^{(i)} \right)\left(m_{t+1}^{(i)} - Am_t^{(i)} \right)^{\top} + S_{t+1}^{(i)} + AS_{t}^{(i)}A^{\top} \right]\boldsymbol{\Sigma}^{-1} \right\},
\end{align*}
where $\mu_{t}^{(i)} := O_t^{(i)}\mathbf{1}_d + \beta_{0} + \beta^\top X_t^{(i)}$ ($\mathbf{1}_d$ denotes a $d$-dimensional vector of ones), $s_t^{(i)}:=(s_{t1}^{(i)},\ldots,s_{td}^{(i)})^\top$,  and   $S_{t}^{(i)} \in \mathbb{R}^{d \times d}$ is the diagonal matrix whose $j$'th diagonal entry is $s_{tj}^{(i)}$.

We estimate $\theta$ and $\Psi$ by maximizing the ELBO via block coordinate ascent. Let $\mathbf{M}$ collect all mean parameters, including the initial-state values, and let $\mathbf{S}$ collect the variational variance parameters for $t\ge 2$. The algorithm cycles through the blocks  $\mathbf{M}, \mathbf{S}, \boldsymbol{\Sigma}^{-1}, A$, and $\left[\beta_0, \beta^{\top} \right]$, updating each one in turn until the increase in ELBO falls below a user-specified tolerance. Although the ELBO is concave in each block separately, it is not jointly concave in all parameters, so convergence to a global maximizer is not guaranteed.

To accommodate estimation in high dimensions, we also develop a penalized VI estimator to encourage estimation of a sparse $A$. Specifically, we first fit the unpenalized VI estimator to obtain pilot estimates $\tilde{A},\tilde{\boldsymbol{\Sigma}}$, compute $\tilde{\boldsymbol{\Sigma}}_Z$, and set $\hat{w}_{jk} = \tilde{\Sigma}^{1/2}_{Z,kk} / \tilde{\Sigma}^{1/2}_{Z,jj}$. We then solve 
$$
\hat{\Psi}_{\mathrm{VI}, \lambda}, \hat{\theta}_{\mathrm{VI}, \lambda} =  \underset{\Psi, \theta}{\arg\min} \quad -\mathrm{ELBO}(y; \Psi, \theta) + \lambda \sum\limits_{j=1}^d\sum\limits_{k=1}^d \hat{w}_{jk}\abs{A_{jk}} . 
$$

To select $\lambda$, the penalized model is fit over a grid of candidate values, and refit on the support selected by $\hat{\theta}_{\mathrm{VI}, \lambda}$, yielding $(\hat{\theta}_{\mathrm{VI, refit}, \lambda},\hat{\Psi}_{\mathrm{VI, refit}, \lambda})$. We then choose $\lambda$ by minimizing the BIC-like criterion: $$
\text{BIC}_{\text{VI}}(\lambda) := -2 \, \mathrm{ELBO}(y; \hat{\Psi}_{\mathrm{VI, refit}, \lambda}, \hat{\theta}_{\mathrm{VI, refit}, \lambda}) + \log (n)\abs{\operatorname{supp}(\hat{A}_{\mathrm{VI, refit},\lambda})}.
$$

%% file: simulations.tex
\section{Simulation Studies}
\label{s:sim}
We conducted simulations to evaluate recovery of the underlying directed networks. Each simulated dataset contained $n\in\{100,250,500\}$ independent trajectories, each observed at $T$ time points. The transition matrix $A$ was sparse with an expected proportion of nonzero entries per row of either $5\%$ or $10\%$. Non-zero entries of $A$ were set to take values uniformly in either  $[0.25, 0.5]$ or $[0.5, 0.75]$. We set $\boldsymbol{\Sigma} = \mathrm{diag}\left(\sigma^2 \right)$, where $\sigma^2 \in \mathbb{R}^d$ and the range of values that entries of $\sigma^2$ could take varied between $[0.1, 0.3]$ or $[0.5, 0.7]$, where each entry sampled uniformly in the given range. Intercepts were generated as $\beta_{0j} \sim N(0.2, 0.1^2)$ and covariate coefficients as $\beta_{kj} \sim N(0, 0.2^2)$.

In each simulation, $Z_1$ was drawn from the stationary distribution of the latent VAR(1) process, and $Z_t$ for $t = 2, \dots, T$ was generated recursively according to Eq.~(\ref{latent_dist}). We generated $p=4$ covariates independently as $X_{tk}^{(i)} \sim N\left(0,2^2 \right)$ for all $t,k$ and $i$. The observed counts were then generated as specified in Eq.~(\ref{obs_cov_dist}).

We first compared how the penalized MoM and VI estimators performed in a low-dimensional setting, fixing $T=5$ and $d=10$. For each setting, we ran 200 simulations and summarized performance by the distributions of the Matthews correlation coefficient (MCC) \citep{matthews1975}. In each run, the penalized MoM estimator was fit over 100 $\lambda_j$ for each row with values ranging from $\lambda_{\max,j}$, the smallest value selecting no edges for row $j$, to $0.01\lambda_{\text{max},j}$. The penalized VI estimator used a common $\lambda$ across all rows, with 100 candidate values ranging from $\lambda_{\max}$, the smallest value selecting no edges in the full network, to $0.01\lambda_{\max}$.

The BIC-selected MoM estimator generally achieved higher MCC than its VI counterpart across the settings considered (Figure~\ref{fig:lowdim_vi_mom_sims}). Recovery generally improved with increasing $n$, with larger improvements for MoM. Based on these results,
we focused the subsequent simulations on the MoM estimator in higher dimensions.

For the higher-dimensional settings, we fixed $T=30$ to reflect the TREAM dataset discussed in Section~\ref{s:realdata}. The sparsity pattern and value ranges of $A$, as well as the value ranges of $\boldsymbol{\Sigma}$ were kept as before.  We considered a medium-dimensional regime, $d = 25$, resembling the TREAM data, and a higher-dimensional regime, $d = 50$, with $n\in \{100,250, 500\}$. For the high-dimensional setting with $d=50$ and values of $A$ in the range $[0.5-0.75]$, we allowed  $5\%$ non-zero values in each row of $A$. This is because we could not reliably construct $A$ matrices with $d=50$ and entries in the $[0.5, 0.75]$ range with all eigenvalues having modulus less than 1 when 10\% of entries in  each row were nonzero. We aimed to assess how model performance changed as dimensionality increases, and also to benchmark our model selection procedure by comparing how the models selected by the BIC-like criteria from Section~\ref{sec:mom_estimator_pen} fared against an oracle estimator, defined as the sparsest model attaining the maximum true positive rate (TPR) along the $\lambda$-grid.  

Here, we present results for $d=25$ with an average of 10\% nonzero entries per row and $d=50$ with an average of 5\% nonzero entries per row in Figures ~\ref{fig:med_dim_sims}--\ref{fig:high_dim_sims}. Network recovery generally improved as $n$ increased in both settings. The BIC-like criterion often achieved TPR close to the oracle benchmark, except when both the transition coefficients $A$, and the latent noise variances $\Sigma$ were relatively small. However, it selected more false edges than the oracle when both were relatively large. For smaller transition coefficients, the oracle-selected networks showed TPR approaching one and false positive rate (FPR) approaching zero as $n$ increased. With larger transition coefficients, the oracle-selected networks continued to exhibit missed and false edges over the sample sizes considered. Additional results for $d=25$ with 5\% density and $d=50$ with 10\% density are shown in Figures~\ref{fig:med_dim_sparse_sims}--\ref{fig:high_dim_dense_sims} in Appendix~\ref{sec:a3}. False positive rates were generally lower in the sparser settings. However, the effect of coefficient magnitude depended on the setting: for $d=25$ with $5\%$ density, larger transition coefficients improved TPR while FPR remained low, with nearly perfect recovery when the latent noise variances were also relatively large.

Larger nonzero transition coefficients might be expected to improve network recovery by increasing signal strength. However, the poorer recovery observed with larger coefficients in some settings may reflect accompanying changes in the stationary latent variances. The covariance of the observed values depends exponentially on the entries of $\boldsymbol{\Sigma}_Z$, which under our model satisfies $\boldsymbol{\Sigma}_Z = A\boldsymbol{\Sigma}_ZA^{\top} + \boldsymbol{\Sigma}$, or equivalently, $\mathrm{vec}\left(\boldsymbol{\Sigma}_Z\right) = (I - A \otimes A)^{-1}\mathrm{vec}(\boldsymbol{\Sigma})$. Hence, larger coefficients in $A$, can lead to larger stationary latent variances, increasing the variability of the observations and empirical moments and making network recovery more difficult in those settings.

%% file: data_analysis.tex
\section{Analysis of Ecological Data }
\label{s:realdata}
We apply our MoM estimator to infer relationships among freshwater macroinvertebrates,  which are often used as bioindicators of aquatic environmental change \citep{holt2011}. Understanding these relationships can help disentangle changes in community composition due to species interactions from those driven by environmental factors. The TREAM dataset compiles macroinvertebrate sampling efforts across 22 European countries spanning 1968–2020 \citep{tream2024}. Our aim is to estimate how the abundance of each macroinvertebrate group depends on the lagged abundances of other groups in the community. Whereas previous analyses of TREAM have focused on temporal trends in community and functional diversity \citep{haase2023, sinclair2024, sexton2024, cortes-guzman2026}, our analysis instead targets the network of temporal relationships among groups.

We focus on a subset of TREAM, comprising 248 freshwater sites in Denmark sampled between 1992 and 2020. Each site was sampled at most once per year, so each year was treated as a time point. Not all sites were sampled every year. We assumed that missingness was independent of the counts conditional on the covariates and offsets, and used only observed consecutive-year pairs to estimate lagged moments. Organisms were recorded at varying taxonomic resolution, and counts were aggregated to the order-or-higher level, referred to in the dataset as ``groups". This yielded 27 groups in total. If a group was never observed at a site, its absence at that site was treated as a structural zero (the group does not inhabit the site). For groups recorded at least once at a site, absence from a sampled site-year was treated as a sampling zero. Years without sampling at the site remained missing. This assumes that the set of groups present at a site is stable over time. We retained only groups that were present in at least 5\% of the available site-year combinations in the dataset and had at least one consecutive-year pair of observations with every other group. This resulted in a final set of 15 groups in our analysis.

For each site-year combination, we included four time-varying covariates: cumulative precipitation in the preceding year (millimeters), the average of monthly maximum temperatures over the preceding 12 months (C\degree), and the upstream percentages of cropland and urban land cover. Four site-level covariates that were constant over time were also included: Strahler stream order (quantifies river branching complexity), flow accumulation (km$^2$, measuring upstream drainage area), elevation (meters), and average stream slope. These covariates were included to control for environmental drivers.

The data were analyzed using the sparse MoM estimator, where due to the low dimensionality of the data we employed exhaustive search to select the final network. As the observed counts for the categories varied greatly in terms of the scale and amount of extreme counts present, we also utilized a stability selection procedure motivated by \citep{meinshausen2010} in our analysis to limit the inclusion of spurious edges in the final network. We implemented this by drawing 200 subsamples consisting of 198 sites (80\% of the total sites) each, fitting the model for each subsample, and designating the stable selected support as edges that were selected in at least 90\% of the subsamples. 
This procedure yielded a final network that selected all self-edges other than that of Heteroptera and 6 cross-edges. In contrast, fitting the model to the full data yielded a much denser network with all of the self-edges and 125 cross-edges selected. Most edges selected from the full data were not consistently selected across subsamples, indicating substantial sensitivity of the estimated support to the sites included in the analysis. After finding the stable support, we performed the refitting procedure to obtain our final estimate of the network, shown in Figure~\ref{fig:tream_network}. 

The recovered network identifies edges among the taxonomic orders of Diptera, Bivalvia, Tricladida, Prosobranchia, Hirudinoidea, Heteroptera, and Odonata. The positive edge from Diptera to Hirudinoidea is compatible with reported feeding relationship between Hirudinoidea and Diptera \citep{toman1997}. The positive edge from Bivalvia to Megaloptera suggests that findings of aquatic mollusks having a positive impact on abundances of other macrofauna from the meta-analysis by \citet{zieritz2025} could also be present in the Danish aquatic sites. We also recovered negative edges from Tricladida to Megaloptera, Heteroptera, and Odonata and a negative edge from Prosobranchia to Odonata. These negative edges may suggest competitive  relationships among these taxonomic orders, but require further ecological investigation. A caveat to these results is that they are based on annual sampling of freshwater sites, so our model may not capture relationships as faithfully if the true lagged effect occurs on a different time scale \citep{tank2019identifiability}.

%% file: discussion.tex
\section{Discussion}
\label{s:discuss}

We proposed a framework for inferring directed networks from multivariate count time series using a latent Gaussian VAR process. To estimate the underlying network, we developed MoM and VI estimators, together with their $\ell_1$-penalized versions for sparse network recovery. We established support recovery and coefficient-consistency results for the penalized MoM estimator under suitable conditions. Our simulation study indicates that the BIC-selected MoM estimator generally outperforms its VI counterpart, and we illustrated the method's utility on freshwater macroinvertebrate abundance data.

Our theoretical network-recovery guarantees require appropriate row-specific penalty sequences. Although the proposed BIC-like criterion performed well in several simulation settings, it omitted many true edges when both the transition coefficients and the latent noise variances were small. The higher sensitivity of the oracle-selected networks in these settings indicates room for improving the tuning
procedure.  
Future work could investigate alternative model-selection criteria or cross-validation procedures based on projected estimating equations, as considered by \citet{xia2026}. Another direction is to extend the proposed estimator to a generalized method-of-moments procedure, weighting the moment conditions according
to their variability and dependence to potentially improve efficiency. This would require reliable estimation of the covariance of the moment conditions. It would also be useful to study an asymptotic regime with a fixed number of trajectories and increasing trajectory lengths, under suitable conditions on temporal dependence, complementing our current theory for many short trajectories. Finally, developing statistical inference procedures for the estimated network coefficients remains an important direction for future work.

\begin{figure}[!h]
\centering
\includegraphics[scale=0.75]{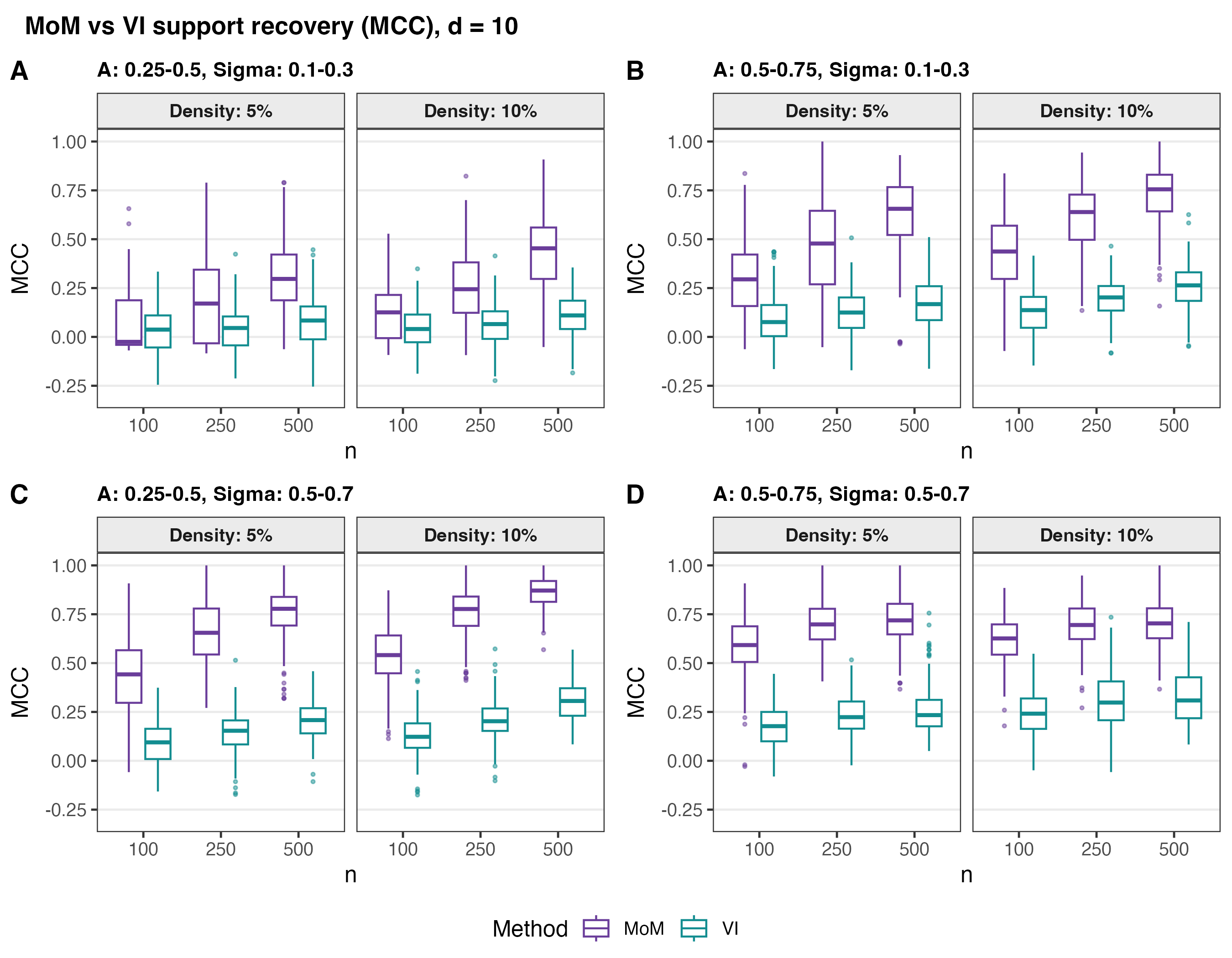}
\caption{Distributions of MCC for the MoM and VI estimators for $d = 10$ across values of $A, \boldsymbol{\Sigma}$, and $n$, where the final model was selected according to the BIC criterion. Within each setting, panels are faceted by the density of $A$. The x-axis gives $n$ and color distinguishes the MoM estimator (purple) from the VI estimator (dark teal).}
\label{fig:lowdim_vi_mom_sims}
\end{figure}

\begin{figure}[!h]
\centering
\includegraphics[scale=0.75]{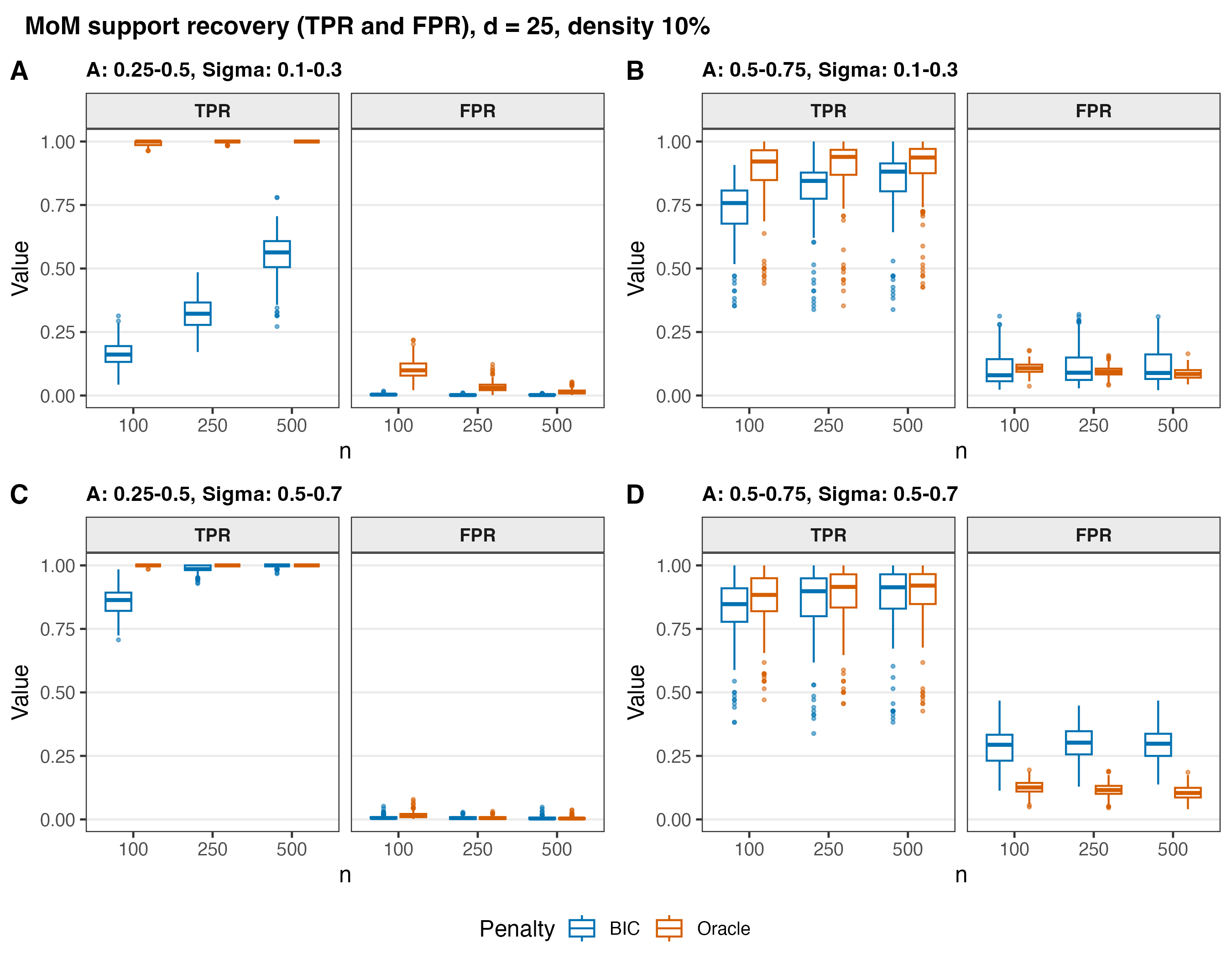}
\caption{Distributions of TPR and FPR for the MoM estimator across values of $A, \boldsymbol{\Sigma}$, and $n$ when $d=25$ and on average 10\% of each row of $A$ is non-zero. Within each setting, panels are faceted by the metric (TPR, FPR). The x-axis gives $n$ and color distinguishes the network selection criteria (``BIC" for BIC and ``Oracle" for the sparsest network attaining the highest TPR), where ``BIC" is in blue and ``Oracle" in orange.}
\label{fig:med_dim_sims}
\end{figure}

\begin{figure}[!h]
\centering
\includegraphics[scale=0.75]{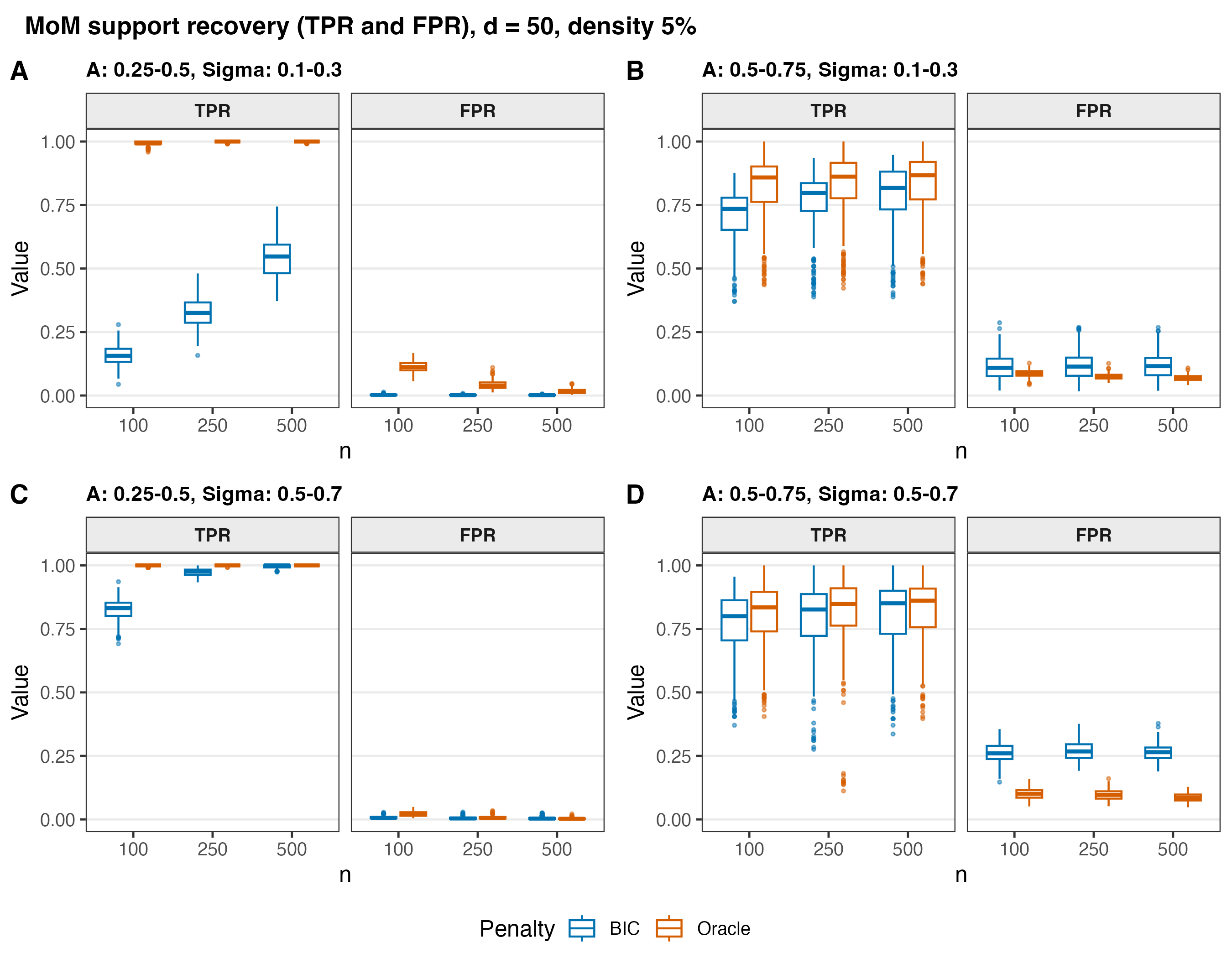}
\caption{Distributions of TPR and FPR for the MoM estimator across values of $A, \boldsymbol{\Sigma}$, and $n$ when $d=50$ and on average 5\% of each row of $A$ is non-zero. Within each setting, panels are faceted by the metric (TPR, FPR). The x-axis gives $n$ and color distinguishes the network selection criteria (``BIC" for BIC and ``Oracle" for the sparsest network attaining the highest TPR), where ``BIC" is in blue and "Oracle" in orange.}
\label{fig:high_dim_sims}
\end{figure}

\begin{figure}[h!]
\centering
\includegraphics[scale=0.2]{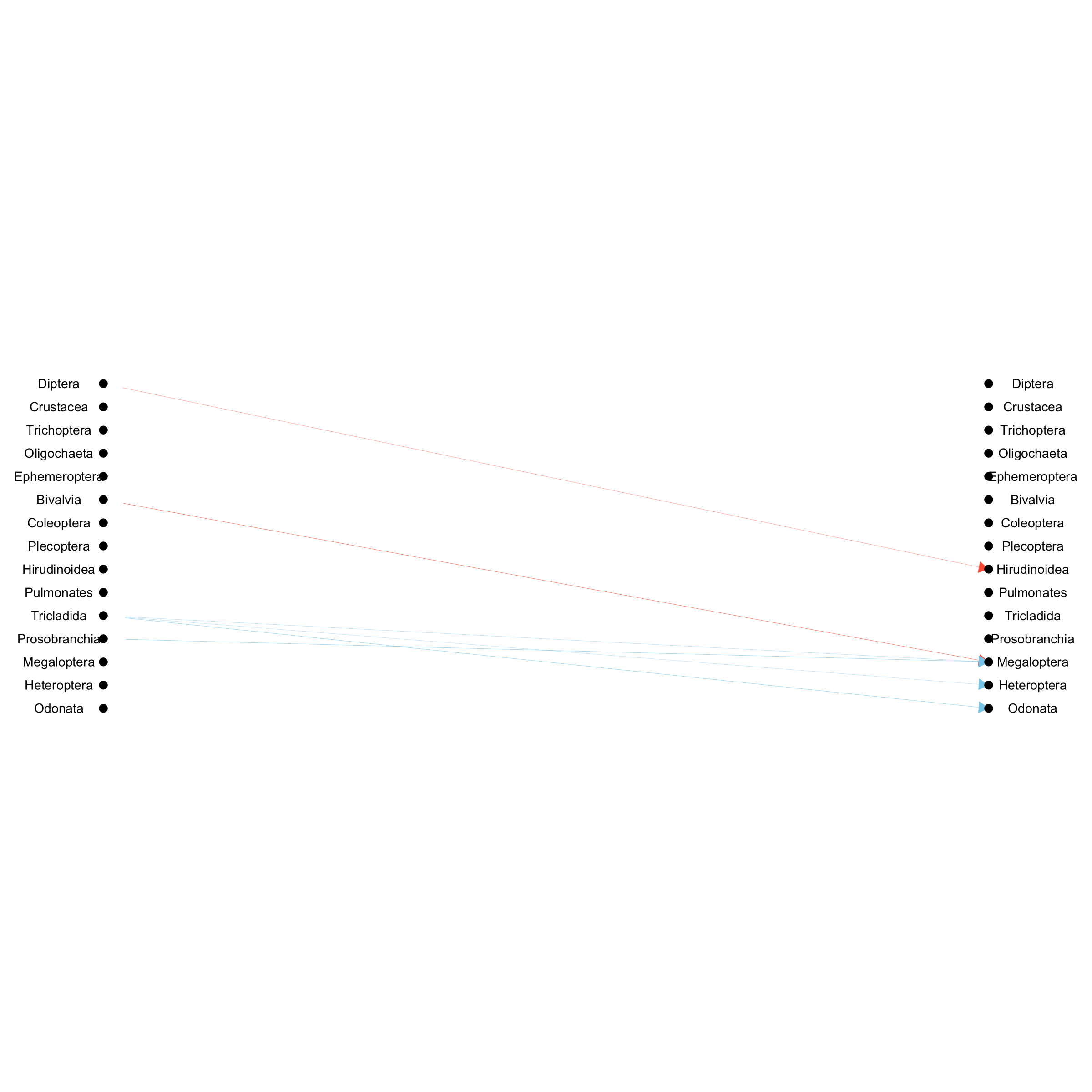}
\caption{Recovered network for the TREAM data, shown as a bipartite graph. Nodes in the left column represent groups at time $t$, and nodes in the right column represent groups at time $t+1$. An edge indicates a lagged effect of a group at time $t$ on a group at time $t+1$. Blue edges indicate negative effects, red edges indicate positive effects, and edge widths are proportional to effect size. Self-edges, corresponding to diagonal entries of $A$, are omitted to highlight interactions among distinct groups; all self-edges except that of Heteroptera were selected and positive.}
\label{fig:tream_network}
\end{figure}

%% file: appendix.tex
\appendix
\setcounter{secnumdepth}{1}
\section{}\label{sec:a1}
\subsection{Appendix A: Proof of Lemma~1: Model Identifiablity} 

\begin{proof}
We establish identifiability of the model by showing that each parameter can be expressed as a function of the moments of $Y_t^{(i)}$. Let $X^{(i)} \in \mathbb{R}^{T_i \times p}$ be the covariates across all time-points for the $i$-th sample and $O^{(i)} \in \mathbb{R}^{T_i}$ be the offsets across all time-points for the $i$-th sample. Then, we have the following expressions for the first and second moments of $Y_{tj}^{(i)}$, the first moment of $Y_{tj}^{(i)}Y_{tk}^{(i)}$, and the first moment of $Y_{t-1,j}^{(i)}Y_{tk}^{(i)}$ , conditional on the covariates, $X^{(i)}$ and offsets $O^{(i)}$.

\begin{align*}
\mathbb{E}\left[Y_{tj}^{(i)} \mid O^{(i)}, X^{(i)} \right] &= \exp\left\{O_{t}^{(i)} + \beta_{0j} + X_{t}^{(i)\top}\beta_j + \frac{1}{2}\boldsymbol{\Sigma}_{Z,jj} \right\}  \\
\mathbb{E}\left[Y_{tj}^{(i)2} \mid O^{(i)}, {X}^{(i)} \right] &= \mathbb{E}\left[Y^{(i)}_{tj} \mid O^{(i)}, {X}^{(i)} \right]\left(1 + \mathbb{E}\left[Y^{(i)}_{tj} \mid O^{(i)}, X^{(i)}\right]\exp\left\{\boldsymbol{\Sigma}_{Z,jj} \right\} \right) \\
\mathbb{E}\left[Y_{tj}^{(i)}Y_{tk}^{(i)} \mid O^{(i)}, X^{(i)} \right] &= \mathbb{E}\left[Y_{tj}^{(i)} \mid O^{(i)}, X^{(i)} \right]\mathbb{E}\left[Y_{tk}^{(i)} \mid O^{(i)}, X^{(i)} \right]\exp\left\{\boldsymbol{\Sigma}_{Z,jk} \right\} \\
\mathbb{E}\left[Y^{(i)}_{tj}Y^{(i)}_{t-1,k} \mid O^{(i)}, X^{(i)} \right] &= \mathbb{E}\left[Y^{(i)}_{tj} \mid O^{(i)}, X^{(i)}\right]\mathbb{E}\left[Y^{(i)}_{t-1,k} \mid O^{(i)}, X^{(i)}\right]\exp\left\{A_j\boldsymbol{\Sigma}_{Z,k} \right\} 
\end{align*} 

From above, we can identify $\boldsymbol{\Sigma}_Z$ by identifying each component of $\boldsymbol{\Sigma}_Z$.

\begin{align*}
\boldsymbol{\Sigma}_{Z,jj} &= \log\left(\mathbb{E}\left[Y_{tj}^{(i)2} \mid O^{(i)}, {X}^{(i)}\right] - \mathbb{E}\left[Y^{(i)}_{tj} \mid O^{(i)}, {X}^{(i)} \right]\right) - 2\log\left(\mathbb{E}\left[Y^{(i)}_{tj} \mid O^{(i)}, {X}^{(i)}\right]\right) \\
\boldsymbol{\Sigma}_{Z,jk} &= \log\left(\mathbb{E}\left[Y^{(i)}_{tj}Y^{(i)}_{tk} \mid O^{(i)}, {X}^{(i)}\right] \right) - \log\left(\mathbb{E}\left[Y^{(i)}_{tj} \mid O^{(i)}, {X}^{(i)} \right] \right) - \log\left(\mathbb{E}\left[Y^{(i)}_{tk} \mid O^{(i)}, {X}^{(i)}\right] \right)
\end{align*}

Let $N_{\mathrm{tot}} := \sum\limits_{i=1}^n T_i$, $\mathbf{X} \in \mathbb{R}^{N_{\mathrm{tot}} \times p}$ be the collection of covariates for each sample and time-point combination, $\mathbf{Y} \in \mathbb{R}^{N_{\mathrm{tot}} \times d}$ be the collection of the observed counts for each category across each time-point and sample combination, and $\mathbf{O} \in \mathbb{R}^{N_{\mathrm{tot}}}$ be the collection of the offset term for each sample and time-point combination. By Lemma~\ref{lem:identification}, $\bar{\mathbf{X}}=[\mathbf{1},\mathbf{X}]$ has full column rank. Since $\Sigma_Z$ has already been identified, the intercept and slopes are identified jointly by
\begin{align*}
\bar{\mathbf{X}}\begin{pmatrix}\beta_{0j}\\\beta_j\end{pmatrix}
&=\log\left(\mathbb{E}[\mathbf{Y}_j\mid\mathbf{O},\mathbf{X}]\right)
-\mathbf{O}-\frac{1}{2}\Sigma_{Z,jj}\mathbf{1},\\
\begin{pmatrix}\beta_{0j}\\\beta_j\end{pmatrix}
&=(\bar{\mathbf{X}}^\top\bar{\mathbf{X}})^{-1}\bar{\mathbf{X}}^\top \left\{\log\left(\mathbb{E}[\mathbf{Y}_j\mid\mathbf{O},\mathbf{X}]\right)
-\mathbf{O}-\frac{1}{2}\Sigma_{Z,jj}\mathbf{1}\right\}.
\end{align*}

The logarithm is taken entrywise. Then, we can identify $A$ by noting that

$$A_j\boldsymbol{\Sigma}_{Z,k} = \log\left(\mathbb{E}\left[Y^{(i)}_{tj}Y^{(i)}_{t-1,k} \mid \mathbf{O}, \mathbf{X}^{(i)} \right] \right) - \log\left(\mathbb{E}\left[Y^{(i)}_{tj} \mid \mathbf{O}, \mathbf{X}^{(i)}\right] \right) - \log\left(\mathbb{E}\left[Y^{(i)}_{t-1,k} \mid \mathbf{O}, \mathbf{X}^{(i)}\right] \right).$$

As we assume that $\boldsymbol{\Sigma}_Z$ is invertible, this implies that
$$ A\boldsymbol{\Sigma}_{Z} = P \iff A = P\boldsymbol{\Sigma}_{Z}^{-1},$$

where 
$$P_{jk} = \log\left(\mathbb{E}\left[Y^{(i)}_{tj}Y^{(i)}_{t-1,k} \mid \mathbf{O}, \mathbf{X}^{(i)} \right] \right) - \log\left(\mathbb{E}\left[Y^{(i)}_{tj} \mid \mathbf{O}, \mathbf{X}^{(i)}\right] \right) - \log\left(\mathbb{E}\left[Y^{(i)}_{t-1,k} \mid \mathbf{O}, \mathbf{X}^{(i)}\right] \right)
$$
Hence, $A$ is identified.

Then, from the stationarity of the latent VAR(1) process, we have that
$$ \boldsymbol{\Sigma} = \boldsymbol{\Sigma}_Z - A\boldsymbol{\Sigma}_ZA^{\top} = \boldsymbol{\Sigma}_Z - P\boldsymbol{\Sigma}_Z^{-1}P^{\top},$$

so $\boldsymbol{\Sigma}$ is identified. Hence, the model is identifiable.
\end{proof}

\section{} \label{sec:a2}
\subsection{Appendix B: Proof of Support Recovery} 
In this appendix, $L_j$ is the empirical Poisson criterion in Section~\ref{sec:mom_estimator}, and
$S^{(j)}=\{k:A_{jk}\neq0\}$, $s_j=\abs{S^{(j)}}$. Individual rows may have empty support. To define the minimum nonzero signal, assume that $A$ has at least one nonzero entry. For a matrix $M$, $M_S$ denotes the columns indexed by $S$ and $S^{(j),\C}$ denotes the complementary set.

We use the augmented design $\bar{\mathbf{X}}$ defined in Lemma~\ref{lem:identification}, with $\bar{X}_t^{(i)}$ denoting the transpose of its row corresponding to trajectory $i$ and time $t$. Let 
$$
r_n=\sqrt{\frac{\log d}{n}}.
$$
Denote $s_n=\max_{1\leq j\leq d}|S^{(j)}|$, $\rho_n = r_n+dr_n^2$,   
and define the minimum nonzero signal by 
$$a_{\min,n}=\min_{\substack{1\leq j,k\leq d\\A_{jk}\neq0}}
|A_{jk}|.$$
For conciseness, define
$$ H_j = \Sigma_{Z, S^{(j)}}^{\top}\Sigma_{Z, S^{(j)}}, \qquad  B_j = \Sigma_{Z,S^{(j),\C}}^\top\Sigma_{Z,S^{(j)}} 
$$
Also define
$$
w_k=\Sigma_{Z,kk}^{1/2},\quad
\hat{w}_k=\hat{\Sigma}_{Z,kk}^{1/2},\quad
D=\operatorname{diag}(w_1,\ldots,w_d),\quad
\hat{D}=\operatorname{diag}(\hat{w}_1,\ldots,\hat{w}_{d}).
$$
$D_S$ and $\hat{D}_S$ denote the diagonal submatrices indexed by $S$. The population and empirical penalty weights are $w_{jk}=w_k/w_j$ and $\hat{w}_{jk}=\hat{w}_k/\hat{w}_j$, respectively. Empirical standard deviations and ratio weights are evaluated on the event $\min_k\hat{\Sigma}_{Z,kk}>0$. Lemma~\ref{lem:gram-weight-rates} shows that this event has probability tending to one.

Assume:
\begin{enumerate}[label=(\Alph*)] 
\item\label{as:trajectories}  The $n$ trajectories are mutually independent. The trajectory lengths are fixed, and there exists a constant $T_{\max}<\infty$, not
depending on $n$, such that
$$
2\leq T_i\leq T_{\max}, \qquad i=1,\ldots,n \, .
$$
\item\label{as:stationarity}
The latent process is stationary and satisfies the model in Section~\ref{sec:model_setup}. For each trajectory, the latent process is independent of the covariates and offsets.

\item\label{as:boundedness} The covariate dimension $p$ is fixed. For every $a>0$, there exists a constant $C_a<\infty$, independent of $n,i,$ and $t$, such that
$$
\E\left\{\exp\left(a\norm{X_t^{(i)}}_2\right)\right\}\leq C_a.
$$
Moreover, there exists a constant $K_O<\infty$ such that
$$
\abs{O_t^{(i)}}\leq K_O \quad\text{almost surely}
$$
for all $i$ and $t$. The true vectors $\gamma_j$ belong to the interior of a fixed compact convex set
$\Gamma\subset\mathbb{R}^{p+1}$ and remain uniformly separated from its boundary.
\item\label{as:gamma_Hessian}
For some constant $c_\gamma>0$,
$$
\inf_{n\ge 1} \min_{1\le j\le d} \inf_{g\in\Gamma} \lambda_{\min} \left\{\E\nabla^2L_j(g)\right\} \ge c_\gamma.
$$

\item\label{as:dimension_growth} There are constants $K_M,C_d<\infty$ and $\kappa>0$ such that
$$
    \|\Sigma_Z\|_{\max}\vee\|P\|_{\max}\leq K_M, \qquad 2\le d\le C_dn^\kappa.
$$

\item\label{sigmaZ_cov_bound} For some constant $C_1>0$, $\norm{\Sigma_Z}_1 \leq C_1$.

\item\label{eigen_min} For some constant $C_2 > 0$, $$\min_{\substack{1\leq j\leq {d}\\s_j>0}}\lambda_{\min}\left(H_j \right) \geq C_2 .$$

\item\label{pop_irrepresentability} For some constant $\eta\in(0,1)$,
$$
\max_{\substack{1\leq j\leq d\\s_j>0}}
\norm{D_{S^{(j),\C}}^{-1}B_jH_j^{-1}D_{S^{(j)}}}_\infty
\leq1-\eta.
$$

\item\label{as:weight_variances} For some constant $c_w>0$,
$$
\min_{1\leq k\leq d}\Sigma_{Z,kk}\geq c_w.
$$

\item\label{A_max_bound} For some constant $C_A>0$,  $\norm{A}_{\max} \le C_A $.
\item\label{lambda_selection} The prescribed positive penalty sequences $\lambda_j$ satisfy
$$
\sqrt{s_n}\max_{1\leq j\leq d}\lambda_j=o(a_{\min,n}),
\qquad
s_n\rho_n=o\left(\min_{1\leq j\leq d}\lambda_j\right).
$$

\end{enumerate}

\begin{proof} [Proof of Lemma~\ref{lem:mom-rates}]
We use the following Nagaev inequality \cite[Corollary~1.8]{nagaev1979large}, see also \cite[Remark~31]{li2024concentration}. 

If $V_1,\ldots,V_n$ are independent, centered random variables satisfying $\E|V_i|^q<\infty$ for every $i$ and some $q>2$, then for every $x>0$, 
\begin{equation} \label{eq:Nagaev}
 \Pr\left(\left|\sum_{i=1}^n V_i\right|>x\right) \leq
 C_q x^{-q}\sum_{i=1}^n\E|V_i|^q +2\exp\left\{-c_q\frac{x^2}{\sum_{i=1}^n\E(V_i^2)}\right\} \, ,
\end{equation}
where $C_q,c_q>0$ depend only on $q$.

Let $\mathcal{I}_n$ index all scalar trajectory-level quantities to which \eqref{eq:Nagaev} is applied in this proof. The particular quantity corresponding to each index is specified when it is introduced below. Since $p$ is fixed and each quantity involves at most two response coordinates, $|\mathcal{I}_n|\le C d^2$ for some constant $C<\infty$.

Suppose that $W_{i\ell}$ is the trajectory-level contribution indexed by $\ell\in\mathcal{I}_n$ and, for some fixed $q>2$,
$$
\sup_{i,\ell}\E|W_{i\ell}|^q\leq C_q.
$$
Applying \eqref{eq:Nagaev} to $V_{i\ell}=W_{i\ell}-\E W_{i\ell}$ and using
$$
\E|V_{i\ell}|^q \le 2^q\E|W_{i\ell}|^q, \qquad 	\sum_{i=1}^n\E V_{i\ell}^2\leq Cn,
$$
gives
\begin{equation} \label{eq:Nagaev-mean}
	\Pr\left(\left|\frac{1}{n}\sum_{i=1}^nV_{i\ell}\right|>x\right) \le C_qn^{1-q}x^{-q}+2\exp(-c_qnx^2).
\end{equation}
Consequently, a union bound and choosing $x=Mr_n$ yield 
\begin{align}
\Pr\left(\max_{\ell\in\mathcal{I}_n}\left|\frac{1}{n}\sum_{i=1}^nV_{i\ell}\right|>Mr_n\right)
	 &\le C_qd^2n^{1-q}(Mr_n)^{-q} +2d^2\exp(-c_qnM^2r_n^2) \nonumber\\
	&\le C_qM^{-q}d^2n^{1-q/2}(\log d)^{-q/2} + 2d^{2-c_qM^2}. \label{eq:Nagaev-union}
\end{align}
Choose and fix an integer $q>4\kappa+2$. The first term in \eqref{eq:Nagaev-union} tends to zero because $d\le C_dn^\kappa$. The second term can be made arbitrarily small by choosing $M$ sufficiently large and, for such a fixed $M$, tends to zero when $d\to\infty$. Thus, any collection of at most order $d^2$ such empirical means is uniformly $O_p(r_n)$.

To apply \eqref{eq:Nagaev-union}, it remains to verify the required $q$'th-moment bound for trajectory-level contributions appearing in the score and in the moment estimators. 

Denote the conditional intensity of the Poisson distribution by 
$$
\Lambda^{(i)}_{tj} = \exp \left\{O_t^{(i)} + \beta_{0j} + X^{(i)\top}_t\beta_j + Z^{(i)}_{tj} \right\} \, .
$$
For every fixed integer $m$, the $m$'th moment of a Poisson variable is a polynomial of degree $m$ in its intensity, consequently,
$$
\E\left[(Y^{(i)}_{tj})^m\mid X_t^{(i)},O_t^{(i)},Z_t^{(i)}\right] \le C_m\left\{1 + (\Lambda^{(i)}_{tj})^m \right\} \, .
$$
Conditional independence of the counts given the latent states gives, for $j\ne k$,
$$
\E\left[(Y^{(i)}_{tj}Y^{(i)}_{tk})^q\mid X_t^{(i)},O_t^{(i)},Z_t^{(i)}\right] \le C_q\left\{1 + (\Lambda^{(i)}_{tj})^q \right\}\left\{1 + (\Lambda^{(i)}_{tk})^q \right\} \, ,
$$
and the same holds at two successive time points. Additionally, 
$$
|(Y^{(i)}_{tj})^2 - Y^{(i)}_{tj}|^q \le C_q \left\{(Y^{(i)}_{tj})^{2q} + (Y^{(i)}_{tj})^q \right\} \, ,
$$
hence
$$\E \left\{|(Y^{(i)}_{tj})^2 - Y^{(i)}_{tj}|^q\mid X_t^{(i)},O_t^{(i)},Z_t^{(i)} \right\} \le C_q\left\{1 +(\Lambda^{(i)}_{tj})^{2q}\right\} \, . $$
For the latent states, $Z_{tj}^{(i)}$ is centered Gaussian with variance at most $K_M$, based on Assumption \ref{as:dimension_growth}. Since $e^{a|z|}\le e^{az}+e^{-az}$, we have 
$$
\E\exp\left\{a|Z_{tj}^{(i)}|\right\}\le
	2\exp\left\{\frac{a^2K_M}{2}\right\} \, .
$$
Hence, by the Cauchy--Schwarz inequality, for $s\in\{t,t-1\}$,
$$
\E\exp\left\{a\left(|Z_{tj}^{(i)}|+|Z_{sk}^{(i)}|\right)\right\}
\le \left[\E e^{2a|Z_{tj}^{(i)}|}\E e^{2a|Z_{sk}^{(i)}|}\right]^{1/2} \le 2e^{2a^2K_M}.
$$

We first apply the concentration bound to the score defining $\hat{\gamma}$. For $r=0,\ldots,p$, define the contribution of trajectory $i$ to coordinate $r$ of the score for $\gamma_j$ by
$$
W_{i,jr}^{\gamma}=\frac{n}{N_{\mathrm{tot}}}\sum_{t=1}^{T_i}\bar{X}_{tr}^{(i)}
\left[\exp\left\{O_t^{(i)}+\bar{X}_t^{(i)\top}\gamma_j\right\}-Y_{tj}^{(i)}\right] \, ,
$$
and let $W_{i,j}^{\gamma}=(W_{i,j0}^{\gamma},\ldots,W_{i,jp}^{\gamma})^\top$. Then, 
$$
\nabla{L}_j(\gamma_j) = \frac1n\sum_{i=1}^n W_{i,j}^{\gamma}.
$$
The definition of $\gamma_j$ and Assumption~\ref{as:stationarity} imply that $\E W_{i,jr}^{\gamma}=0$. Compactness of $\Gamma$ bounds the coefficients in $\gamma_j$ uniformly in $j$. Since $\beta_{0j}=\gamma_{0j}-\Sigma_{Z,jj}/2$, the coefficients in the intensity are also uniformly bounded. The conditional Poisson moment bound, compactness of $\Gamma$, and boundedness of the offsets give the following bound for the conditional $q$'th moment of each score summand:
$$
C_q\norm{\bar{X}_t^{(i)}}_2^q \exp\left\{C_q\norm{\bar{X}_t^{(i)}}_2+q|Z_{tj}^{(i)}|\right\}.
$$
Its expectation is uniformly bounded by Cauchy--Schwarz, Assumption~\ref{as:boundedness}, and the Gaussian exponential-moment bound above. Since $T_i\le T_{\max}$ and $N_{\mathrm{tot}}\ge 2n$, the moment bounds above and Jensen's inequality give
$$
\sup_{i,j,r}\E|W_{i,jr}^{\gamma}|^q\le C_q.
$$
There are $d(p+1)$ score coordinates and $p$ is fixed. Therefore, \eqref{eq:Nagaev-union} gives 
\begin{equation} \label{eq:score-rate}
\max_{1\leq j\leq d}\norm{\nabla L_j(\gamma_j)}_2 = O_p(r_n). 	
\end{equation}
The Hessian is 
$$
\nabla^2{L}_j(g) = \frac{1}{N_{\mathrm{tot}}}\sum_{i=1}^n\sum_{t=1}^{T_i}  \exp\left\{O_t^{(i)}+\bar{X}_t^{(i)\top}g\right\}\bar{X}_t^{(i)}\bar{X}_t^{(i)\top}.
$$
Differentiating the Hessian with respect to coordinate $g_r$ gives
$$
\frac{\partial}{\partial g_r}\nabla^2 L_j(g)
=\frac{1}{N_{\mathrm{tot}}}\sum_{i=1}^n\sum_{t=1}^{T_i}
\exp\left\{O_t^{(i)}+\bar{X}_t^{(i)\top}g\right\}
\bar{X}_{tr}^{(i)}\bar{X}_t^{(i)}\bar{X}_t^{(i)\top}.
$$
Since $\Gamma$ is compact and the offsets are bounded, the absolute values of the scalar summands in the Hessian and its derivatives are bounded, uniformly over $g\in\Gamma$, by
$$
C\left(1+\norm{\bar{X}_t^{(i)}}_2^3\right) \exp\left\{C\norm{\bar{X}_t^{(i)}}_2\right\}.
$$
Assumption~\ref{as:boundedness} gives uniformly bounded $q$'th moments for these bounds. Their empirical averages are $O_p(1)$ by Markov's inequality, and their expectations are uniformly bounded. Compactness gives a finite net of $\Gamma$ at any fixed approximation accuracy, with cardinality independent of $n,d$. The number of scalar Hessian entries evaluated at its points is therefore at most $Cd^2$, with $C$ allowed to depend on the net. Applying \eqref{eq:Nagaev-union} at these points gives an $o_p(1)$ bound there. The derivative bounds control the difference between any point and a net point. Taking increasingly fine nets after $n\to\infty$ therefore gives
$$
\max_{1\leq j\leq d}\sup_{g\in\Gamma}\norm{ \nabla^2{L}_j(g)-\E\nabla^2{L}_j(g)}_2 = o_p(1).
$$

For any $j$ and $g\in\Gamma$,
$$
\lambda_{\min}\left\{\nabla^2{L}_j(g)\right\} \ge \lambda_{\min}\left\{\E\nabla^2{L}_j(g)\right\} - \norm{\nabla^2{L}_j(g)-\E\nabla^2{L}_j(g)}_{2}
$$
Therefore, Assumption~\ref{as:gamma_Hessian} together with the preceding uniform $o_p(1)$ bound imply that, with probability tending to one,
\begin{equation} \label{eq:empirical-strong-convexity}
\min_{1\leq j\leq d}\inf_{g\in\Gamma} \lambda_{\min}\left\{\nabla^2{L}_j(g)\right\} \ge c_\gamma/2.
\end{equation}
Uniform separation from the boundary of $\Gamma$ gives a fixed $\epsilon>0$ such that the ball of radius $\epsilon$ around each $\gamma_j$ lies in $\Gamma$. By \eqref{eq:empirical-strong-convexity}, on its boundary, 
$$
L_j(g)-L_j(\gamma_j) \geq-\epsilon\norm{\nabla L_j(\gamma_j)}_2 +\frac{c_\gamma}{4}\epsilon^2.
$$
The score bound \eqref{eq:score-rate} makes this positive simultaneously for all $j$ with probability tending to one. Thus, each ball has an interior minimizer. Convexity of $L_j$ and strict convexity within $\Gamma$ make it the unique global minimizer $\hat{\gamma}_j$. Consequently,
$$
\Pr\left(\hat{\gamma}_j\in\Gamma \text{ for every }1\leq j\leq d \right) \longrightarrow 1.
$$

On this event, strong convexity and the minimizing property of $\hat{\gamma}_j$ imply
$$
0 \ge {L}_j(\hat{\gamma}_j)-{L}_j(\gamma_j) \ge \nabla{L}_j(\gamma_j)^\top (\hat{\gamma}_j-\gamma_j) +\frac{c_\gamma}{4} \norm{\hat{\gamma}_j-\gamma_j}_2^2.
$$
Consequently,
$$
\norm{\hat{\gamma}_j-\gamma_j}_2 \le \frac{4}{c_\gamma}\norm{\nabla{L}_j(\gamma_j)}_2,
$$
and \eqref{eq:score-rate} yields 
\begin{equation} \label{eq:gamma-rate}
\max_{1\leq j\leq d}\norm{\hat{\gamma}_j-\gamma_j}_2=O_p(r_n).
\end{equation}
We now apply the same concentration result to the moment estimators. Define the trajectory-level contributions evaluated at the true $\gamma$ by the following expressions.  The first is for $j\neq k$, whereas the lagged expression is for all $j,k$,
\begin{align*}
W_{i,jk}^{\Sigma} &= \frac{n}{N_{\mathrm{tot}}} \sum_{t=1}^{T_i} Y_{tj}^{(i)}Y_{tk}^{(i)} \exp\left\{-2O_t^{(i)} -\bar{X}_t^{(i)\top}\gamma_j -\bar{X}_t^{(i)\top}\gamma_k \right\},\\
W_{i,jj}^{\Sigma} &= \frac{n}{N_{\mathrm{tot}}} \sum_{t=1}^{T_i} \left\{(Y_{tj}^{(i)})^2-Y_{tj}^{(i)}\right\} \exp\left\{-2O_t^{(i)}-2\bar{X}_t^{(i)\top}\gamma_j \right\},\\
W_{i,jk}^{P} &= \frac{n}{N_{\mathrm{tot}}-n}\sum_{t=2}^{T_i}Y_{tj}^{(i)}Y_{t-1,k}^{(i)}\exp\left\{-O_t^{(i)}-O_{t-1}^{(i)}-\bar{X}_t^{(i)\top}\gamma_j-\bar{X}_{t-1}^{(i)\top}\gamma_k\right\}.
\end{align*}
We now verify the moment bound for these specific summands. For example, for $j\neq k$, the conditional $q$'th moment of the summand in $W_{i,jk}^{\Sigma}$ satisfies
\begin{align*}
&\E\left[\left|Y_{tj}^{(i)}Y_{tk}^{(i)}
e^{-2O_t^{(i)}-\bar{X}_t^{(i)\top}\gamma_j-\bar{X}_t^{(i)\top}\gamma_k}\right|^q
\mid X_t^{(i)},O_t^{(i)},Z_t^{(i)}\right]\\
&\qquad\leq C_q e^{-q\{2O_t^{(i)}+\bar{X}_t^{(i)\top}(\gamma_j+\gamma_k)\}} 	\left\{1+(\Lambda_{tj}^{(i)})^q\right\}
\left\{1+(\Lambda_{tk}^{(i)})^q\right\}.
\end{align*}
Expanding the right-hand side gives finitely many exponential terms with bounded covariate coefficients and fixed multiples of the latent coordinates. Their expectations are uniformly finite by Assumption~\ref{as:boundedness}, the Gaussian exponential-moment bounds, and Cauchy--Schwarz inequality. The same calculation, using the factorial-moment bound above when $j=k$ and the two-time-point bound for the lagged products, gives a uniformly bounded $q$'th moment for every time-point summand in $W_{i,jk}^{\Sigma}$ and $W_{i,jk}^{P}$.

Since $T_i\le T_{\max}$, Jensen's inequality gives
$$
\left|\sum_{t=1}^{T_i} a_t\right|^q \le T_i^{q-1}\sum_{t=1}^{T_i}|a_t|^q \le T_{\max}^{q-1}\sum_{t=1}^{T_i}|a_t|^q.
$$
Moreover, $N_{\mathrm{tot}}\ge 2n$, so the factors $n/N_{\mathrm{tot}}$ and $n/(N_{\mathrm{tot}}-n)$ in the definitions of $W_{i,jk}^{\Sigma}$ and $W_{i,jk}^{P}$ are uniformly bounded. Therefore,
$$
\sup_{i,j,k}\E|W_{i,jk}^{\Sigma}|^q\le C_q, \qquad \sup_{i,j,k}\E|W_{i,jk}^{P}|^q\leq C_q.
$$

The oracle empirical moments, obtained by using the true $\gamma$ in place of $\hat{\gamma}$, can therefore be written as
$$
\tilde{M}_{jk}^{\Sigma} = \frac{1}{n}\sum_{i=1}^n W_{i,jk}^{\Sigma}, \qquad  \tilde{M}_{jk}^{P} = \frac{1}{n}\sum_{i=1}^n W_{i,jk}^{P} \, .
$$
The moments identities established in Appendix \ref{sec:a1} imply that, with $M^{\Sigma}_{jk}=e^{\Sigma_{Z,jk}}$ and $M^{P}_{jk}=e^{P_{jk}}$,
$$
\frac{1}{n}\sum_{i=1}^n\E W_{i,jk}^{\Sigma}=M_{jk}^{\Sigma}, \qquad \frac{1}{n}\sum_{i=1}^n\E W_{i,jk}^{P}=M_{jk}^{P} \, .
$$
It follows directly that
$$
\tilde{M}_{jk}^{\Sigma}-M_{jk}^{\Sigma} = \frac{1}{n}\sum_{i=1}^n \left(W_{i,jk}^{\Sigma}-\E W_{i,jk}^{\Sigma}\right), \qquad
\tilde{M}_{jk}^{P}-M_{jk}^{P} = \frac{1}{n}\sum_{i=1}^n \left(W_{i,jk}^{P}-\E W_{i,jk}^{P}\right) \, .
$$

There are at most order $d^2$ such moment entries. Applying \eqref{eq:Nagaev-union} to the preceding centered representations therefore gives
\begin{equation} \label{eq:oracle-moment-rate}
    \norm{\tilde{M}^{\Sigma}-M^{\Sigma}}_{\max}=O_p(r_n), \qquad \norm{\tilde{M}^{P}-M^{P}}_{\max}=O_p(r_n).
\end{equation}
It remains to account for replacing $\gamma$ by $\hat{\gamma}$ in the empirical moments. For $j\neq k$, consider the empirical moment as a function of two arguments:
$$
\hat{M}_{jk}^{\Sigma}(g,h):= \frac{1}{N_{\mathrm{tot}}} \sum_{i=1}^n\sum_{t=1}^{T_i} Y_{tj}^{(i)}Y_{tk}^{(i)} \exp\left\{-2O_t^{(i)}-\bar{X}_t^{(i)\top}g
-\bar{X}_t^{(i)\top}h\right\}.
$$
Then $\tilde{M}_{jk}^{\Sigma}$ is obtained by evaluating this function at $(\gamma_j,\gamma_k)$, whereas $\hat{M}_{jk}^{\Sigma}$ is obtained by evaluating it at $(\hat{\gamma}_j,\hat{\gamma}_k)$.

Differentiation with respect to either $g$ or $h$ multiplies each summand by $-\bar{X}_t^{(i)}$. Compactness of $\Gamma$ and boundedness of the offsets therefore give
\begin{align*}
	&\sup_{g,h\in\Gamma}\left\{
	\norm{\nabla_g\hat{M}_{jk}^{\Sigma}(g,h)}_2+
	\norm{\nabla_h\hat{M}_{jk}^{\Sigma}(g,h)}_2\right\}\\
	&\leq\frac{2}{N_{\mathrm{tot}}}
	\sum_{i=1}^n\sum_{t=1}^{T_i}
	\norm{\bar{X}_t^{(i)}}_2Y_{tj}^{(i)}Y_{tk}^{(i)}
	\exp\left\{2K_O+C\norm{\bar{X}_t^{(i)}}_2\right\}.
\end{align*}
The extra covariate factor and the exponential factor are also controlled by Assumption~\ref{as:boundedness}. The conditional Poisson bounds and Cauchy Schwartz inequality thus give bounded $q$'th moments for the trajectory contributions on the right, by the same calculation used above. Their expectations are uniformly bounded, and \eqref{eq:Nagaev-union} bounds their centered averages uniformly over $j\neq k$. Consequently,
$$
\max_{j\neq k}\sup_{g,h\in\Gamma}\left\{
\norm{\nabla_g\hat{M}_{jk}^{\Sigma}(g,h)}_2+
\norm{\nabla_h\hat{M}_{jk}^{\Sigma}(g,h)}_2\right\}=O_p(1).
$$
On the high probability event established above, both $\gamma_j$ and $\hat{\gamma}_j$ belong to $\Gamma$ for every $j$. Since $\Gamma$ is convex, the line segment joining them is also contained in $\Gamma$. The mean-value theorem then gives, uniformly over $j\neq k$,
$$
\left|\hat{M}_{jk}^{\Sigma}-\tilde{M}_{jk}^{\Sigma}\right| \le O_p(1) \left( \norm{\hat{\gamma}_j-\gamma_j}_2 + \norm{\hat{\gamma}_k-\gamma_k}_2\right).
$$
When $j=k$, the corresponding empirical moment is
$$
\hat{M}_{jj}^{\Sigma}(g):= \frac{1}{N_{\mathrm{tot}}} \sum_{i=1}^n\sum_{t=1}^{T_i} \left\{ \left(Y_{tj}^{(i)}\right)^2-Y_{tj}^{(i)}\right\}\exp\left\{-2O_t^{(i)}
-2\bar{X}_t^{(i)\top}g\right\}.
$$
Its derivative multiplies each summand by $-2\bar{X}_t^{(i)}$. The same argument therefore gives
$$
\max_{1\leq j\leq d} \left|\hat{M}_{jj}^{\Sigma}-\tilde{M}_{jj}^{\Sigma}\right|\le O_p(1)\max_{1\leq j\leq d}\norm{\hat{\gamma}_j-\gamma_j}_2.
$$
For the lagged moment, the corresponding function is
$$
\hat{M}_{jk}^{P}(g,h):= \frac{1}{N_{\mathrm{tot}}-n}\sum_{i=1}^n\sum_{t=2}^{T_i}Y_{tj}^{(i)}Y_{t-1,k}^{(i)}\exp\left\{-O_t^{(i)}-O_{t-1}^{(i)}-\bar{X}_t^{(i)\top}g-\bar{X}_{t-1}^{(i)\top}h\right\}.
$$
Its derivatives multiply the summands by $-\bar{X}_t^{(i)}$ and $-\bar{X}_{t-1}^{(i)}$, respectively. The same moment and concentration bounds give
$$
\max_{1\leq j,k\leq d}\left|\hat{M}_{jk}^{P}-\tilde{M}_{jk}^{P}\right|\le O_p(1)\max_{1\leq j\leq d}\norm{\hat{\gamma}_j-\gamma_j}_2.
$$
Using Eq.~\eqref{eq:gamma-rate}, we conclude that
$$
\norm{\hat{M}^{\Sigma}-\tilde{M}^{\Sigma}}_{\max} = O_p(r_n),\qquad \norm{\hat{M}^{P}-\tilde{M}^{P}}_{\max} = O_p(r_n).
$$
Combining these bounds with \eqref{eq:oracle-moment-rate} gives
$$
\norm{\hat{M}^{\Sigma}-M^{\Sigma}}_{\max}=O_p(r_n),
\qquad \norm{\hat{M}^{P}-M^{P}}_{\max}=O_p(r_n).
$$
Assumption~\ref{as:dimension_growth} gives
$M_{jk}^{\Sigma}\geq e^{-K_M}$ and $M_{jk}^{P}\geq e^{-K_M}$.
Since $r_n\to0$, all empirical moments are at least $e^{-K_M}/2$ with probability tending to one. On this event the logarithm is uniformly Lipschitz. Applying it entrywise proves the two remaining rates in Lemma~\ref{lem:mom-rates}. The estimators are thus well defined on an event whose probability tends to one.

\end{proof}

\begin{lemma}\label{lem:gram-weight-rates}
Let $\hat{H}_j=\hat{\Sigma}_{Z,S^{(j)}}^\top
\hat{\Sigma}_{Z,S^{(j)}}$.
Under Assumptions~\ref{as:trajectories}--\ref{eigen_min} and~\ref{as:weight_variances}--\ref{lambda_selection},
uniformly over rows with $s_j>0$,
$$
\norm{\hat{H}_j-H_j}_{\max}=O_p(\rho_n),
\qquad
\norm{\hat{H}_j^{-1}}_\infty=O_p(\sqrt{s_j}).
$$
Moreover, all estimated diagonal entries are positive with
probability tending to one, and the penalty weights satisfy
$$
\max_{j,k}\abs{\hat{w}_{jk}-w_{jk}}=O_p(r_n).
$$
\end{lemma}

\begin{proof}[Proof of Lemma~\ref{lem:gram-weight-rates}]

Writing  $\hat{\Delta}_{\Sigma_Z} = \hat{\Sigma}_Z - \Sigma_Z,$ we have
\begin{align*}
\norm{\hat{H}_j - H_j}_{\max} &=  \norm{\hat{\Delta}^{\top}_{\Sigma_{Z},S^{(j)}}\hat{\Delta}_{\Sigma_{Z},S^{(j)}} + \Sigma^{\top}_{Z,S^{(j)}}\hat{\Delta}_{\Sigma_{Z},S^{(j)}} + \hat{\Delta}^{\top}_{\Sigma_{Z},S^{(j)}}\Sigma_{Z,S^{(j)}}}_{\max} \\
&\leq \norm{\hat{\Delta}_{\Sigma_{Z},S^{(j)}}}_1\norm{\hat{\Delta}_{\Sigma_{Z},S^{(j)}}}_{\max}+2\norm{\Sigma_{Z,S^{(j)}}^\top}_\infty \norm{\hat{\Delta}_{\Sigma_{Z},S^{(j)}}}_{\max} \\
&\leq
d\norm{\hat{\Delta}_{\Sigma_Z}}^2_{\max}+2\norm{\Sigma_Z}_{1}\norm{\hat{\Delta}_{\Sigma_Z}}_{\max} \\
&\leq
d\norm{\hat{\Delta}_{\Sigma_Z}}^2_{\max}+2 C_1\norm{\hat{\Delta}_{\Sigma_Z}}_{\max} \, .
\end{align*}
Combining the preceding bound with Lemma~\ref{lem:mom-rates},
uniformly over $j$,
\begin{equation*} 
  \norm{\hat{H}_j - H_j}_{\max} = O_p(dr_n^2 + r_n)=O_p(\rho_n).  
\end{equation*}
Now, we provide a lower bound for $\lambda_{\min}\left(\hat{H}_j \right)$.
\begin{align*}
\lambda_{\min}\left(\hat{H}_j \right) &= \lambda_{\min}\left(\hat{H}_j - H_j + H_j \right) \\
&\geq \lambda_{\min}\left(H_j \right)  + \lambda_{\min}\left(\hat{H}_j - H_j \right) \\
&\geq  C_2 - \norm{\hat{H}_j - H_j}_2 \\
&\geq C_2 - s_j\norm{\hat{H}_j - H_j}_{\max} \, ,
\end{align*}
Since $s_j\le s_n$, Assumption \ref{lambda_selection} gives $s_j\rho_n\le s_n\rho_n=o(\min_j\lambda_j)$. Also, $\max_j\lambda_j=o(1)$, because $s_n\geq 1$ and $a_{\min,n}\leq C_A$. Hence $s_n\rho_n=o(1)$, so with probability tending to one $\lambda_{\min}(\hat{H}_j)>C_2/2$ simultaneously for all nonempty supports. Consequently,
\begin{equation} \label{eq:Hj_inv_infnorm}
  \norm{\hat{H}_j^{-1}}_\infty \le \sqrt{s_j}\norm{\hat{H}_j^{-1}}_2 = \frac{\sqrt{s_j}}{\lambda_{\min}(\hat{H}_j)} = O_p(\sqrt{s_j})  
\end{equation}
By Assumption~\ref{as:weight_variances} and Lemma~\ref{lem:mom-rates}, all estimated diagonal entries are at least $c_w/2$ with probability tending to one. The square-root and inverse-square-root functions are uniformly Lipschitz on this interval. Consequently,
$$
\max_k\abs{\hat{w}_k-w_k}=O_p(r_n),
\qquad
\max_k\abs{\hat{w}_k^{-1}-w_k^{-1}}=O_p(r_n).
$$ 
Since $\hat{w}_{jk}=\hat{w}_k/\hat{w}_j$,
$$
\abs{\hat{w}_{jk}-w_{jk}} \leq \hat{w}_j^{-1}\abs{\hat{w}_k-w_k} +w_k\abs{\hat{w}_j^{-1}-w_j^{-1}}.
$$
Assumption~\ref{as:dimension_growth} gives $w_k\leq\sqrt{K_M}$.
On the event that all estimated diagonal entries are at least
$c_w/2$, we also have $\hat{w}_j^{-1}\leq\sqrt{2/c_w}$.
Therefore,
$$
\max_{j,k}\abs{\hat{w}_{jk}-w_{jk}}=O_p(r_n).
$$
\end{proof}

We are now ready to prove Theorem~\ref{thm:support-recovery}.

\begin{proof} [Proof of Theorem~\ref{thm:support-recovery}]
The problem we wish to solve is 

\begin{equation*}
\underset{A \in \mathbb{R}^{d \times d}}{\arg\min} \norm{A\hat{\Sigma}_Z - \hat{P}}_F^2 + \sum\limits_{j=1}^d\lambda_j\sum\limits_{k = 1}^d \hat{w}_{jk}\abs{A_{jk}}
\end{equation*}

Let $a^{(j)}$, $ p^{(j)}, $ and $ \hat{p}^{(j)} $ denote the corresponding rows as column vectors. Since $\hat{\Sigma}_Z$ is symmetric, the problem is equivalent to minimizing
\begin{equation*}
\sum\limits_{j=1}^d\left\{\norm{\hat{\Sigma}_Za^{(j)} - \hat{p}^{(j)}}_2^2 + \frac{\lambda_j}{\hat{w}_{j}} \norm{\hat{D}a^{(j)}}_1\right\} \, .
\end{equation*}

We work on the high-probability event in Lemma~\ref{lem:gram-weight-rates}. Let $S^{(j)} = \left\{k : A_{jk} \neq 0\right\} $ denote the true support of $a^{(j)}$. For each $j$, construct a candidate solution by solving the problem restricted to $S^{(j)}$:
$$\tilde{a}^{(j)}_{S^{(j)}} \in
\underset{a_{S^{(j)}}\in\mathbb{R}^{|S^{(j)}|}}{\arg\min}
\left\|
\hat\Sigma_{Z,S^{(j)}}a_{S^{(j)}}-\hat p^{(j)}
\right\|_2^2
+
\frac{\lambda_j}{\hat{w}_j}\|\hat{D}_{S^{(j)}}a_{S^{(j)}}\|_1 \, ,$$   
and set $\tilde{a}_{S^{(j),\mathcal{C}}}^{(j)}=0$.

Then, for any $j$, the KKT condition for the restricted problem implies that there exists 
$$\tilde{z}^{(j)}_{S^{(j)}} \in \partial \left\|\tilde{a}^{(j)}_{S^{(j)}}\right\|_1, \qquad \left\|\tilde{z}^{(j)}_{S^{(j)}}\right\|_\infty \le 1 ,$$
such that
\begin{equation} \label{eq:subgrad}
2\hat{\Sigma}_{Z, S^{(j)}}^{\top}\left[\hat{\Sigma}_{Z, S^{(j)}}\tilde{a}_{S^{(j)}}^{(j)} - \hat{p}^{(j)} \right] + \frac{\lambda_j}{\hat{w}_j}\hat{D}_{S^{(j)}} \tilde{z}^{(j)}_{S^{(j)}}  = 0.    
\end{equation}

Solving for $\tilde{a}_{S^{(j)}}^{(j)}$, yields
$$ \tilde{a}_{S^{(j)}}^{(j)} = \hat{H}_j^{-1}\left[\hat{\Sigma}_{Z, S^{(j)}}^{\top}\hat{p}^{(j)} - \frac{\lambda_j}{2\hat{w}_j}\hat{D}_{S^{(j)}} \tilde{z}^{(j)}_{S^{(j)}}\right]$$
Since $p^{(j)}=\Sigma_{Z,S^{(j)}}a^{(j)}_{S^{(j)}}$, and letting $\hat{\Delta}_{p^{(j)}} = \hat{p}^{(j)}-p^{(j)}$,
we obtain
\begin{equation} \label{eq:kkt_active}
\tilde{a}_{S^{(j)}}^{(j)} - a_{S^{(j)}}^{(j)} =\hat{H}_j^{-1}\left[\hat{\Sigma}_{Z, S^{(j)}}^{\top}\left(\hat{\Delta}_{p^{(j)}} - \hat{\Delta}_{\Sigma_Z,S^{(j)}} a_{S^{(j)}}^{(j)}  \right) - \frac{\lambda_j}{2\hat{w}_j}\hat{D}_{S^{(j)}} \tilde{z}^{(j)}_{S^{(j)}}  \right].     
\end{equation}

Therefore,
$$\norm{\tilde{a}_{S^{(j)}}^{(j)} - a_{S^{(j)}}^{(j)}}_\infty \le \norm{\hat{H}_j^{-1}}_\infty \left[\norm{\hat{\Sigma}_{Z, S^{(j)}}}_1\left(\norm{\hat{\Delta}_{p^{(j)}}}_\infty + \norm{\hat{\Delta}_{\Sigma_Z,S^{(j)}}}_{\max} \norm{a_{S^{(j)}}^{(j)}}_1  \right) + \frac{\lambda_j}{2\hat{w}_j}\|\hat{D}_{S^{(j)}}\|_\infty\right] .$$
From Lemma~\eqref{lem:mom-rates} and Assumption~\ref{sigmaZ_cov_bound}, uniformly over $j$, 
\begin{equation} \label{eq:norm1hatSigmaZ}
\norm{\hat{\Sigma}_{Z, S^{(j)}}}_1\le \norm{\Sigma_Z}_1 + d\norm{\hat{\Sigma}_Z - \Sigma_Z}_{\max} = O_p(1 + dr_n) \, ,  
\end{equation}
and from Assumption \ref{A_max_bound} we have $\norm{a_{S^{(j)}}^{(j)}}_1\le C_As_j$. Together with Lemmas~\eqref{lem:mom-rates}--\eqref{lem:gram-weight-rates}, and Eq.~\eqref{eq:Hj_inv_infnorm}, these bounds yield 
$$\norm{\tilde{a}_{S^{(j)}}^{(j)} - a_{S^{(j)}}^{(j)}}_\infty =O_p(s_j^{3/2}\rho_n+\sqrt{s_j}\lambda_j) .$$
Consequently,
$$\max_{1\leq j\leq d}\norm{\tilde{a}_{S^{(j)}}^{(j)} - a_{S^{(j)}}^{(j)}}_\infty =O_p(s_n^{3/2}\rho_n+\sqrt{s_n}\max_j\lambda_j) .$$
By Assumption \ref{lambda_selection},
$$
\frac{{s_n}^{3/2}\rho_n}{a_{\min,n}} \leq\frac{{s_n}\rho_n}{\min_j\lambda_j}
\frac{\sqrt{{s_n}}\max_j\lambda_j}{a_{\min,n}}\longrightarrow0.
$$
Thus, $s_n^{3/2}\rho_n+\sqrt{s_n}\max_j\lambda_j = o(a_{\min,n})$, so
it follows that
$$\max_{1\leq j\leq d}\norm{\tilde{a}_{S^{(j)}}^{(j)} - a_{S^{(j)}}^{(j)}}_\infty =o_p(a_{\min,n}) .$$
Therefore,
$$\Pr\left(\max_{1\leq j\leq d}\norm{\tilde{a}_{S^{(j)}}^{(j)} - a_{S^{(j)}}^{(j)}}_\infty < a_{\min,n} \right)\longrightarrow 1 \, .$$
On this event, for every $j$ and every $k\in S^{(j)}$,
$$\left| \tilde{a}_k^{(j)} - a_k^{(j)}\right|<a_{\min,n}\le \left|a_k^{(j)} \right| .$$
Hence, $\tilde{a}_k^{(j)}\neq 0$, and $\operatorname{sign}(\tilde{a}_k^{(j)})=\operatorname{sign}(a_k^{(j)})$.
Thus, with probability tending to one, the restricted  candidate has no false negatives and recovers the signs of all coefficients on the true support. Since $\tilde{a}_{S^{(j),\mathcal{C}}}^{(j)} =0$, by construction, it follows that $\operatorname{supp}(\tilde{a}^{(j)})=S^{(j)}$, uniformly over $j$, with probability tending to one. This establishes support and sign recovery for the restricted candidate. To show that this candidate is a solution to the full optimization problem, it remains to construct the inactive subgradient and verify that 
$$\norm{\tilde{z}^{(j)}_{S^{(j),\mathcal{C}}}}_\infty<1. $$
For convenience, let us define 
$$\hat{B}_j = \hat{\Sigma}_{Z,S^{(j),\C}}^\top\hat{\Sigma}_{Z,S^{(j)}}    , \qquad R_j = \hat{p}^{(j)} - \hat{\Sigma}_{Z,S^{(j)}}a_{S^{(j)}}^{(j)} $$
Define the inactive subgradient by the full KKT equation:
\begin{equation*}
\tilde{z}_{S^{(j),\C}}^{(j)} =-\frac{2\hat{w}_j}{\lambda_j}\hat{D}_{S^{(j),\C}}^{-1} \hat{\Sigma}_{Z,S^{(j),\C}}^\top (\hat{\Sigma}_{Z,S^{(j)}}\tilde{a}_{S^{(j)}}^{(j)}-\hat{p}^{(j)}).
\end{equation*}
Combining this with \eqref{eq:kkt_active} gives
\begin{align}\label{eq:ztilde_substituted}
\tilde{z}_{S^{(j),\C}}^{(j)}
&=\hat{D}_{S^{(j),\C}}^{-1}\hat{B}_j\hat{H}_j^{-1} \hat{D}_{S^{(j)}}\tilde{z}_{S^{(j)}}^{(j)}\nonumber\\
&+\frac{2\hat{w}_j}{\lambda_j}\hat{D}_{S^{(j),\C}}^{-1}
\left(\hat{\Sigma}_{Z,S^{(j),\C}}^\top -\hat{B}_j\hat{H}_j^{-1}\hat{\Sigma}_{Z,S^{(j)}}^\top\right)R_j.
\end{align}

We would like now to transfer Assumption \ref{pop_irrepresentability} to its sample counterpart. For any $k\in S^{(j),\C}$ and $\ell\in S^{(j)}$, and from Eq.~\eqref{eq:norm1hatSigmaZ},
\begin{align*}
    \left|\hat{\Sigma}_{Z,k}^\top\hat{\Sigma}_{Z,\ell} - \Sigma_{Z,k}^\top\Sigma_{Z,\ell}\right| &\leq \norm{\hat{\Sigma}_{Z,k}-\Sigma_{Z,k}}_\infty \norm{\hat{\Sigma}_{Z,\ell}}_1 + \norm{\Sigma_{Z,k}}_1 \norm{\hat{\Sigma}_{Z,\ell}-\Sigma_{Z,\ell}}_\infty \\
   &= O_p\left\{r_n(1+dr_n) + r_n \right\} = O_p(\rho_n) \, ,
\end{align*}
uniformly in $j,k$ and $\ell$. Since $\hat{B}_j - B_j$ has $s_j$ columns, it follows that 
$$\norm{\hat{B}_j - B_j}_\infty = O_p(s_j\rho_n) \, ,$$
uniformly in $j$. Using the identity
$$\hat{B}_j\hat{H}^{-1}_j - B_jH^{-1}_j = (\hat{B}_j - B_j)\hat{H}_j^{-1} + B_jH_j^{-1}(H_j - \hat{H}_j)\hat{H}_j^{-1} \, .$$
The population weights and their inverses are uniformly bounded. Thus Assumption~\ref{pop_irrepresentability} implies $\norm{B_jH_j^{-1}}_\infty=O(1)$ uniformly in $j$. Together with the preceding bounds, we obtain
$$\norm{\hat{B}_j\hat{H}^{-1}_j - B_jH^{-1}_j}_\infty \leq \norm{\hat{B}_j - B_j}_\infty\norm{\hat{H}_j^{-1}}_\infty + \norm{B_jH_j^{-1}}_\infty\norm{H_j - \hat{H}_j}_\infty\norm{\hat{H}_j^{-1}}_\infty = O_p\left(s_j^{3/2}\rho_n\right) \, .$$
Consequently,
\begin{equation}\label{eq:BH-rate}
\max_{1\leq j\leq d}\norm{\hat{B}_j\hat{H}^{-1}_j - B_jH^{-1}_j}_\infty = O_p\left(s_n^{3/2}\rho_n\right) = o_p(1) \, .
\end{equation}
We now compare the sample weighted matrix with its population counterpart. Combining \eqref{eq:BH-rate} with the bounds for $\hat{D}$ and $\hat{D}^{-1}$ established in the proof of Lemma~\ref{lem:gram-weight-rates} gives
\begin{align*}
\max_{1\leq j\leq d}\big\|\hat{D}_{S^{(j),\C}}^{-1}\hat{B}_j\hat{H}_j^{-1}\hat{D}_{S^{(j)}}-D_{S^{(j),\C}}^{-1}B_jH_j^{-1}D_{S^{(j)}}\big\|_\infty
=O_p({\color{blue}s_n}^{3/2}\rho_n+r_n)=o_p(1).
\end{align*}
Consequently, with probability tending to one,
$$
\max_{1\leq j\leq d} \norm{\hat{D}_{S^{(j),\C}}^{-1}\hat{B}_j\hat{H}_j^{-1}\hat{D}_{S^{(j)}}}_\infty \leq1-\frac{\eta}{2}.
$$
Since $\norm{R_j}_\infty = O_p(s_jr_n)$, uniformly in $j$, and 
$$\norm{\hat{\Sigma}_{Z,S^{(j),\C}}}_1 + \norm{\hat{\Sigma}_{Z,S^{(j)}}}_1 = O_p(1+dr_n) \, ,$$
we have that, uniformly in $j$,
\begin{align*}
\norm{\left(\hat{\Sigma}_{Z,S^{(j),\C}}^\top -\hat{B}_j\hat{H}_j^{-1}\hat{\Sigma}_{Z,S^{(j)}}^\top\right)R_j}_\infty &\leq 
\left\{\norm{\hat{\Sigma}_{Z,S^{(j),\C}}}_1 + \norm{\hat{B}_j\hat{H}_j^{-1}}_\infty \norm{\hat{\Sigma}_{Z,S^{(j)}}}_1\right\} \norm{R_j}_\infty \\
&= O_p\left\{s_jr_n(1+dr_n)\right\} = O_p(s_j\rho_n) \, .
\end{align*}
Because $\left\|\tilde{z}^{(j)}_{S^{(j)}}\right\|_\infty \le 1$, and the standard deviations and their inverses are uniformly bounded, together with \eqref{eq:ztilde_substituted} and Assumption~\ref{lambda_selection} give
$$
\max_{1\leq j\leq d}\left\|\tilde{z}^{(j)}_{S^{(j),\C}}\right\|_\infty \le 1-\frac{\eta}{2} + O_p\left(\frac{s_n\rho_n}{\min_j\lambda_j}\right) = 1-\frac{\eta}{2}+o_p(1)  \, .
$$
Thus strict dual feasibility holds simultaneously for all rows, and each restricted candidate is a minimizer of the full rowwise problem. It remains to prove uniqueness. Let $\bar{a}^{(j)}$ be any other minimizer. Expanding the difference of the objective values and using $2\hat{\Sigma}_Z^\top(\hat{\Sigma}_Z\tilde{a}^{(j)}-\hat{p}^{(j)})
+\frac{\lambda_j}{\hat{w}_j}\hat{D}\tilde{z}^{(j)}=0$ gives
\begin{equation*}
    0=\norm{\hat{\Sigma}_Z(\bar{a}^{(j)}-\tilde{a}^{(j)})}_2^2+\lambda_j\sum_{k=1}^{d}\hat{w}_{jk}
\left\{\abs{\bar{a}_k^{(j)}}-\abs{\tilde{a}_k^{(j)}}-\tilde{z}_k^{(j)}(\bar{a}_k^{(j)}-\tilde{a}_k^{(j)})\right\}.
\end{equation*}
Both terms, and each summand in the second term, are nonnegative. The contribution of the inactive coordinates to the second term is at least
$$
\lambda_j\sum_{k\in S^{(j),\C}}\hat{w}_{jk}
\left(1-\abs{\tilde{z}_k^{(j)}}\right)\abs{\bar{a}_k^{(j)}}.
$$
The weights are positive and strict dual feasibility holds, so $\bar{a}_{S^{(j),\C}}^{(j)}=0$. The first term also vanishes, giving
$$
\hat{\Sigma}_{Z,S^{(j)}}(\bar{a}_{S^{(j)}}^{(j)}-\tilde{a}_{S^{(j)}}^{(j)})=0.
$$
Positive definiteness of $\hat{H}_j$ implies equality of the active coefficients. Thus the minimizer is unique. Combining this with the established support and sign recovery proves
$$
\Pr\left\{\operatorname{supp}(\hat{A})=\operatorname{supp}(A),
\ \operatorname{sign}(\hat{A})=\operatorname{sign}(A)\right\}\longrightarrow1.
$$
\end{proof}

\begin{proof} [Proof of Corollary~\ref{cor:coefficient-consistency}]
On the support-recovery event, $\hat{A}$ equals the restricted candidate constructed in the proof of Theorem~\ref{thm:support-recovery}. The coefficient-error bound established there gives the first rate.

On the same event, the refit uses the true support in every row. For rows with nonempty support, its normal equations give
$$
\hat{a}_{S^{(j)}}^{(j),\mathrm{refit}}-a_{S^{(j)}}^{(j)} = \hat{H}_j^{-1}\hat{\Sigma}_{Z,S^{(j)}}^\top R_j.
$$
Using \eqref{eq:Hj_inv_infnorm}, \eqref{eq:norm1hatSigmaZ}, and the previously established bound $\norm{R_j}_\infty=O_p(s_jr_n)$ yields, uniformly over these rows,
$$
\norm{\hat{a}_{S^{(j)}}^{(j),\mathrm{refit}} -a_{S^{(j)}}^{(j)}}_\infty = O_p\left(s_j^{3/2}r_n(1+dr_n)\right) = O_p(s_j^{3/2}\rho_n).
$$
Rows with empty support are refitted as zero. Since the support-recovery event has probability tending to one, taking the maximum over rows gives the asserted refit rate.

Finally, Assumption~\ref{lambda_selection} implies
$$
s_n^{3/2}\rho_n=o(a_{\min,n}), \qquad \sqrt{s_n}\max_j\lambda_j=o(a_{\min,n}).
$$
Since $a_{\min,n}\leq C_A$ by Assumption~\ref{A_max_bound}, both estimators are consistent in the maximum-entry norm.
\end{proof}

\section{}\label{sec:a3}
\subsection{Appendix C: Additional simulation results}

Here, we provide additional simulation results for the settings where $d=25$ and on average 5\% of each row of $A$ is non-zero and for the settings where $d=50$ and on average 10\% of each row of $A$ is non-zero.

\begin{figure}[!h]
\centering
\includegraphics[scale=0.75]{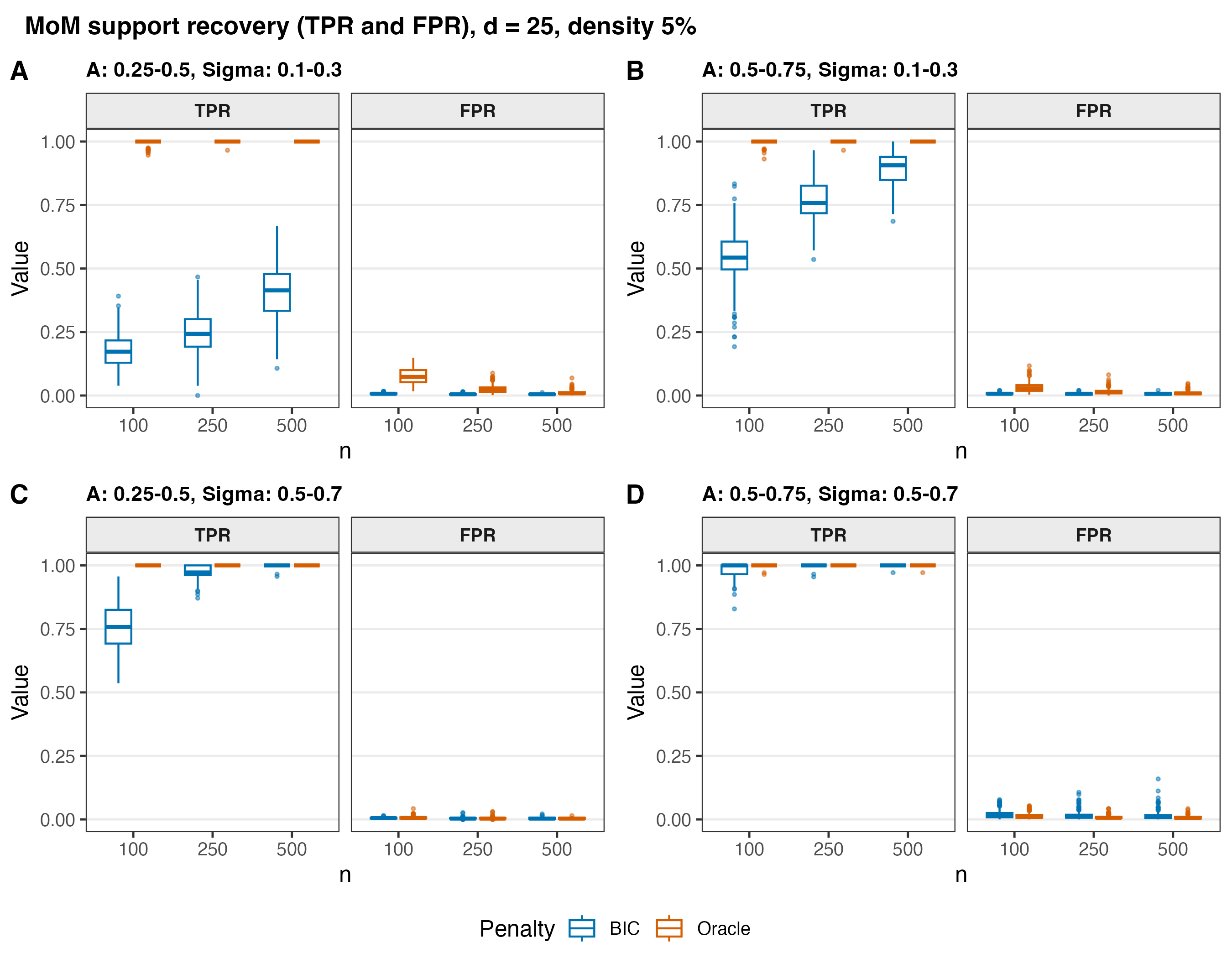}
\setcounter{figure}{0}
\gdef\thefigure{A\arabic{figure}}
\gdef\theHfigure{appendix.A\arabic{figure}}
\caption{Distributions of TPR and FPR for the MoM estimator across values of $A, \boldsymbol{\Sigma}$, and $n$ when $d=25$ and on average 5\% of each row of $A$ is non-zero. Within each setting, panels are faceted by the metric (TPR, FPR). The x-axis gives $n$ and color distinguishes the network selection criteria (``BIC" for BIC and ``Oracle" for the sparsest network attaining the highest TPR), where ``BIC" is in blue and "Oracle" in orange.}
\label{fig:med_dim_sparse_sims}
\end{figure}

\begin{figure}[!h]
\centering
\includegraphics[scale=0.75]{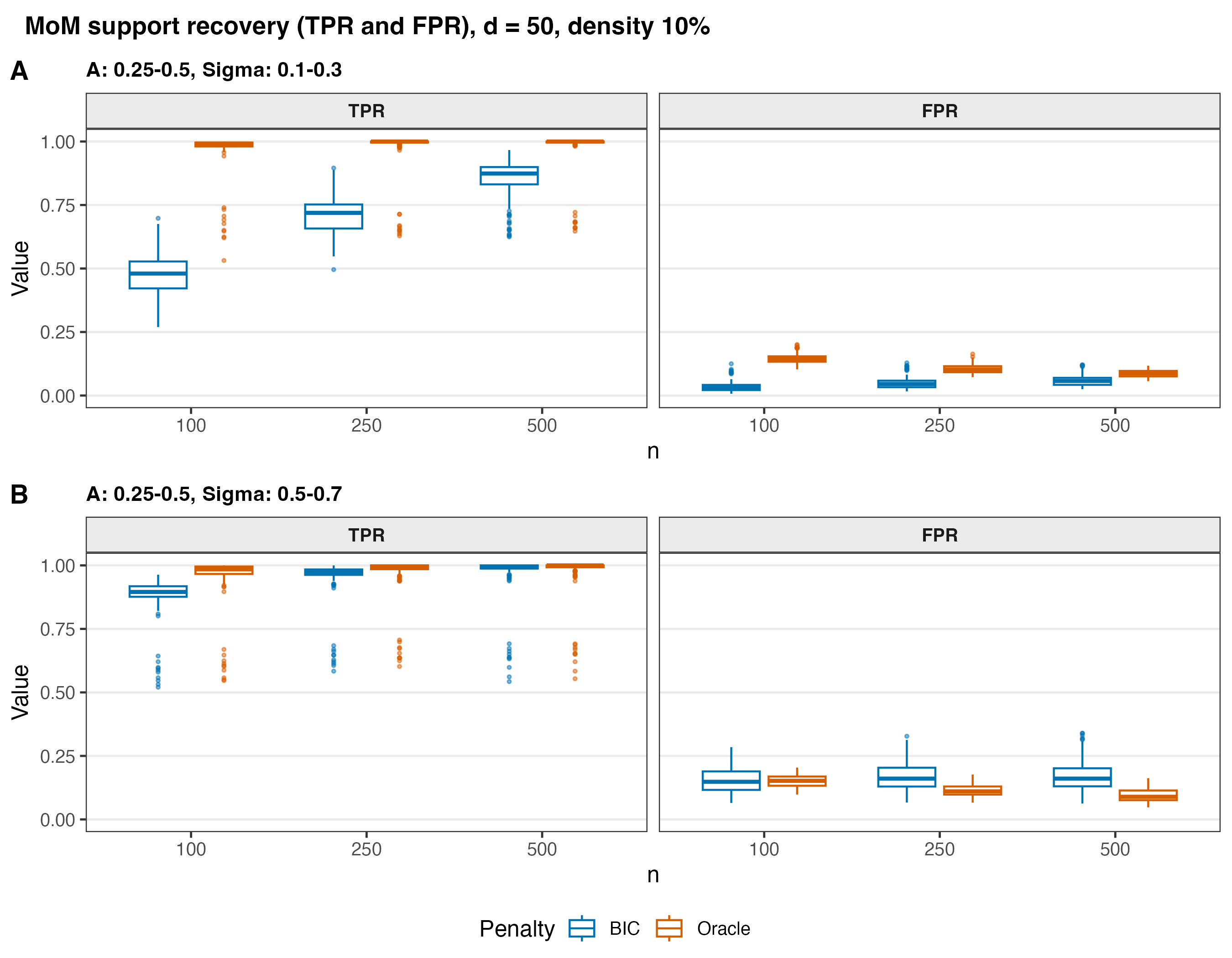}
\caption{Distributions of TPR and FPR for the MoM estimator across values of $A, \boldsymbol{\Sigma}$, and $n$ when $d=50$ and on average 10\% of each row of $A$ is non-zero. Within each setting, panels are faceted by the metric (TPR, FPR). The x-axis gives $n$ and color distinguishes the network selection criteria (``BIC" for BIC and ``Oracle" for the sparsest network attaining the highest TPR), where ``BIC" is in blue and "Oracle" in orange.}
\label{fig:high_dim_dense_sims}
\end{figure}